\documentclass[sigconf,balance=false]{acmart}

\usepackage[utf8]{inputenc}
\usepackage{array,booktabs,arydshln}
\usepackage{tabularx}
\usepackage{colortbl}
\usepackage{amsfonts}
\usepackage{amsmath}
\usepackage{caption}
\usepackage{subcaption}
\usepackage{xcolor}
\usepackage{xurl}
\usepackage{xspace}
\usepackage{float}
\usepackage{graphicx}
\usepackage{adjustbox}
\usepackage{paralist}

\newtheorem{thm}{Theorem}[section]

\newtheorem{lem}[thm]{Lemma}

\newenvironment{proofof}[1]{\indent{\scshape Proof of #1}:~~}{\vspace{3mm}\qed}

\newcommand{\mypara}[1]{\vspace{1pt}\noindent{\bf {#1}}}

\setcopyright{none}
\copyrightyear{YYYY}

\acmYear{YYYY}
\acmVolume{YYYY}
\acmNumber{X}
\acmDOI{XXXXXXX.XXXXXXX}
\acmISBN{}
\acmConference{arXiv}

\renewcommand\footnotetextcopyrightpermission[1]{}

\def\Kern#1{\kern #1\relax}
\begin{document}
\settopmatter{printfolios=true}
\title[DeepSE-WF: Unified Security Estimation for Website Fingerprinting Defenses]{DeepSE-WF: Unified Security Estimation for Website Fingerprinting Defenses}


\author{Alexander Veicht}
\orcid{1234-5678-9012}
\affiliation{%
  \institution{ETH Zurich}
  \city{Zurich}
  \country{Switzerland}}
\email{alexander.veicht@inf.ethz.ch}

\author{Cedric Renggli}
\affiliation{%
  \institution{University of Zurich}
  \city{Zurich}
  \country{Switzerland}}
\email{renggli@ifi.uzh.ch}

\author{Diogo Barradas}
\affiliation{%
  \institution{University of Waterloo}
  \city{Waterloo}
  \state{ON}
  \country{Canada}
}
\email{diogo.barradas@uwaterloo.ca}



\begin{abstract}
  Website fingerprinting (WF) attacks, usually conducted with the help of a machine learning-based classifier, enable a network eavesdropper to pinpoint which website a user is accessing through the inspection of traffic patterns. These attacks have been shown to succeed even when users browse the Internet through encrypted tunnels, e.g., through Tor or VPNs. To assess the security of new defenses against WF attacks, recent works have proposed feature-dependent theoretical frameworks that estimate the Bayes error of an adversary's features set or the mutual information leaked by manually-crafted features. Unfortunately, as WF attacks increasingly rely on deep learning and latent feature spaces, our experiments show that security estimations based on simpler (and less informative) manually-crafted features can no longer be trusted to assess the potential success of a WF adversary in defeating such defenses. 
  In this work, we propose DeepSE-WF, a novel WF security estimation framework that leverages specialized kNN-based estimators to produce Bayes error and mutual information estimates from learned latent feature spaces, thus bridging the gap between current WF attacks and security estimation methods. Our evaluation reveals that DeepSE-WF produces tighter security estimates than previous frameworks, reducing the required computational resources to output security estimations by one order of magnitude.
\end{abstract}


\maketitle

\section{Introduction}

The simple activity of web browsing can pose a threat to users' privacy. Regardless of whether encryption is used to obscure the content of communications, a network eavesdropper may still be able to infer meaningful privacy-sensitive data (like a user's health condition or financial situation) by identifying the website (or sequences of websites) that a user is accessing~\cite{statisticalIdentification}. The problem arises from the fact that encryption hides a communication's contents but not its metadata (e.g. the source and destination of traffic produced by users).

To thwart an eavesdropper's ability to infer privacy-sensitive information resulting from the analysis of browsing metadata, low-latency anonymous communication tools, such as Tor~\cite{Tor}, are able to obscure both the content and destination of communications by routing encrypted traffic through a number of network nodes. However, Tor does not significantly modify the shape of traffic patterns, preserving the packet timing and volume characteristics which are tied to a given website~\cite{lowcost} (mostly to support an interactive browsing experience). Unfortunately, this leaves Tor vulnerable to
two important classes of traffic analysis attacks, commonly denominated by end-to-end confirmation attacks and \textit{website fingerprinting} (WF). The former aim to correlate network flows entering and exiting the Tor network in order to link users with their destinations~\cite{deepcorr,deepCoffea}. The latter are the focus of this paper, and aim to unveil specific websites visited by users through the analysis of Tor connections' traffic characteristics~\cite{knnAttack,Tik-Tok}.

As a response to WF attacks, the research community developed multiple defenses~\cite{tamaraw,wtfpad,front} that aim to reshape a website's original traffic patterns. By adding and/or delaying a connection's packets, these defenses make it hard for an adversary to pinpoint, with high confidence, the website browsed by a given user. Thus, a critical aspect in the evaluation of such defenses is the assessment of their effectiveness against increasingly sophisticated attacks.

Up until recently, the security of WF defenses could only be assessed by testing their ability to defend users from the strongest existing WF attacks. However, this methodology perpetuated an arms race that shed few light on the ability of a defense to survive against an attack specially devised to undermine it. The quest for devising a general security evaluation procedure has led to the surfacing of two main frameworks, one based on the estimation of the Bayes error~\cite{wfes}, and the other one using the concept of mutual information~\cite{wefde}. These frameworks use a set of features, manually crafted by domain experts, to estimate the security of a defense irrespective of the choice of attack or classifier, shifting the focus of WF research to the search for more informative features~\cite{wfes,wefde}.

This quest for more informative features as a means to improve the success of WF attacks led researchers to explore the use of deep neural networks (DNNs)~\cite{AWF,Tik-Tok}. One notorious benefit of this approach is that neural networks are able to automatically extract information-rich latent feature spaces, yielding the most accurate WF attacks to date.
Unfortunately, these lines of work led to a severe dissociation between the (latent) features used to perform state-of-the-art attacks and the manually crafted features used by security estimation frameworks to assess the security of emerging defenses. We observed that these sets of manual features lack the ability to characterize traffic patterns to the same degree of detail obtained by latent feature spaces, and thus that security estimations based on manually-engineered features are no longer reliable indicators of the security of novel defenses against DNN-based attacks.
Further, DNN-based attacks use large datasets to learn meaningful latent features. This means that security estimation frameworks must be able to scale to large numbers of samples to exploit their theoretical behavior and efficiently assess defenses' security guarantees.

To address the above limitations, this paper introduces DeepSE-WF \textit{(\underline{Deep} \underline{S}ecurity \underline{E}stimations for \underline{W}ebsite \underline{F}ingerprinting)}, a novel framework for security estimation of WF defenses. DeepSE-WF overcomes two core limitations of existing security estimation methods.
First, it takes advantage of DNNs to ingest a large number of website traces and build highly-informative latent feature spaces. The features obtained by such a process deliver a more comprehensive characterization of complex traffic interactions when compared to manual feature engineering. 
Second, DeepSE-WF leverages two highly scalable kNN-based estimators which make use of the previously obtained latent feature spaces to compute the Bayes error and mutual information estimates of the security of WF defenses. This effectively bridges the gap between the representations of features used by WF attacks and defense estimation methodologies. Furthermore, by deploying both a Bayes error- and MI-based defense estimator on the same learned latent feature space, our framework \textit{unifies} existing security estimation methods in a single framework. 

DeepSE-WF relies on techniques developed by the fields of machine learning and information theory to estimate both quantities of interest (i.e., the Bayes error and mutual information), and to understand how they are related~\cite{renggli2021evaluating, ross2014mutual, fano1961transmission, hellmanraviv}.
Yet, the impact of deploying these estimators \textit{after} applying a potentially learned feature transformation on top of raw (network traffic) data is largely under-explored. We build on top of one of the few works in this area~\cite{knnConvergence} to motivate DeepSE-WF from a theoretical point of view.

\mypara{Contributions.} We summarize our key contributions as follows:
\begin{compactitem}
    \item We identify a disconnect between the information-rich features used in current WF attacks (obtained from learned latent feature spaces)  and those used by existing WF security estimation approaches (obtained through manual feature engineering). We show that this disconnect can lead to severe overestimates of a WF defense's security guarantees.
    \item We propose a new WF security estimation framework, DeepSE-WF, which takes advantage of learned latent feature spaces to jointly estimate the Bayes error and mutual information achieved by existing WF defenses, and motivate the use of these estimators from a theoretical perspective.
    \item We experimentally evaluate DeepSE-WF and show that it can achieve significantly tighter bounds than previous WF security estimation methods based on the Bayes error (WFES~\cite{wfes}) or mutual information (WeFDE~\cite{wefde}). Our results also reveal that DeepSE-WF is able to scale up the number of samples used in security estimations by 2 orders-of-magnitude vs. WFES, and is up to 7 times faster than WeFDE in producing an estimation. We have publicly released our codebase~\cite{ourCode}.
\end{compactitem}

\section{Website Fingerprinting}
\label{sec:wf_background}

Website fingerprinting (WF) is a class of statistical traffic analysis attacks aimed at identifying which websites a user is visiting, even when encrypted tunnels are used to obscure a user's intended destination (e.g., when browsing through a VPN, or when forwarding traffic through an anonymity network like Tor). These attacks can be launched by passive network adversaries (e.g., an ISP), being imperceptible to users, and rely on the analysis of different indicators obtained from the traffic patterns of a network connection, such as packet direction, timing, or bursting behavior. Since these attacks require only the analysis of network metadata, they can be applied without the need for breaking the cryptographic primitives used to encrypt users' traffic.

The threat model of a WF attack is illustrated in Figure~\ref{fig:wf_attack_diagram}. For launching a WF attack, an adversary typically operates as follows. First, the adversary accesses a pool of websites of interest (monitored pages) $W = \{w_1, w_2, ... , w_n\}$, collecting multiple network traffic observations of accesses to these websites. Following the definition of Cai et al.~\cite{tamaraw}, these observations can be represented as website traces, i.e., packet sequences T=$\langle$(\textit{ipd}$_{1}$,\textit{l}$_{1}$),(\textit{ipd}$_{2}$,\textit{l}$_{2}$),...,(\textit{ipd}$_{n}$,\textit{l}$_{n}$)$\rangle$ where \textit{ipd} represents the inter-packet timing difference between packets \textit{i} and \textit{i-1}, and \textit{l} represents the packet length which can further be signed as a positive value if the packet is outgoing or as a negative value if the packet is incoming. (Note that while WF attacks can make use of packet size information in general settings, this information is usually disregarded when considering the Tor scenario since Tor leverages fixed-sized cells as its basic communication data unit~\cite{cumul}.) Then, the adversary extracts a number of features from these traces (i.e., the fingerprints), and resorts to machine learning (ML) techniques to build a model which, given any user trace $t$, predicts the corresponding visited website $\hat{w_t} \in W$. WF attacks can be typically launched in two different settings:

\begin{figure}[t]
    \centering
        \includegraphics[width=0.85\columnwidth]{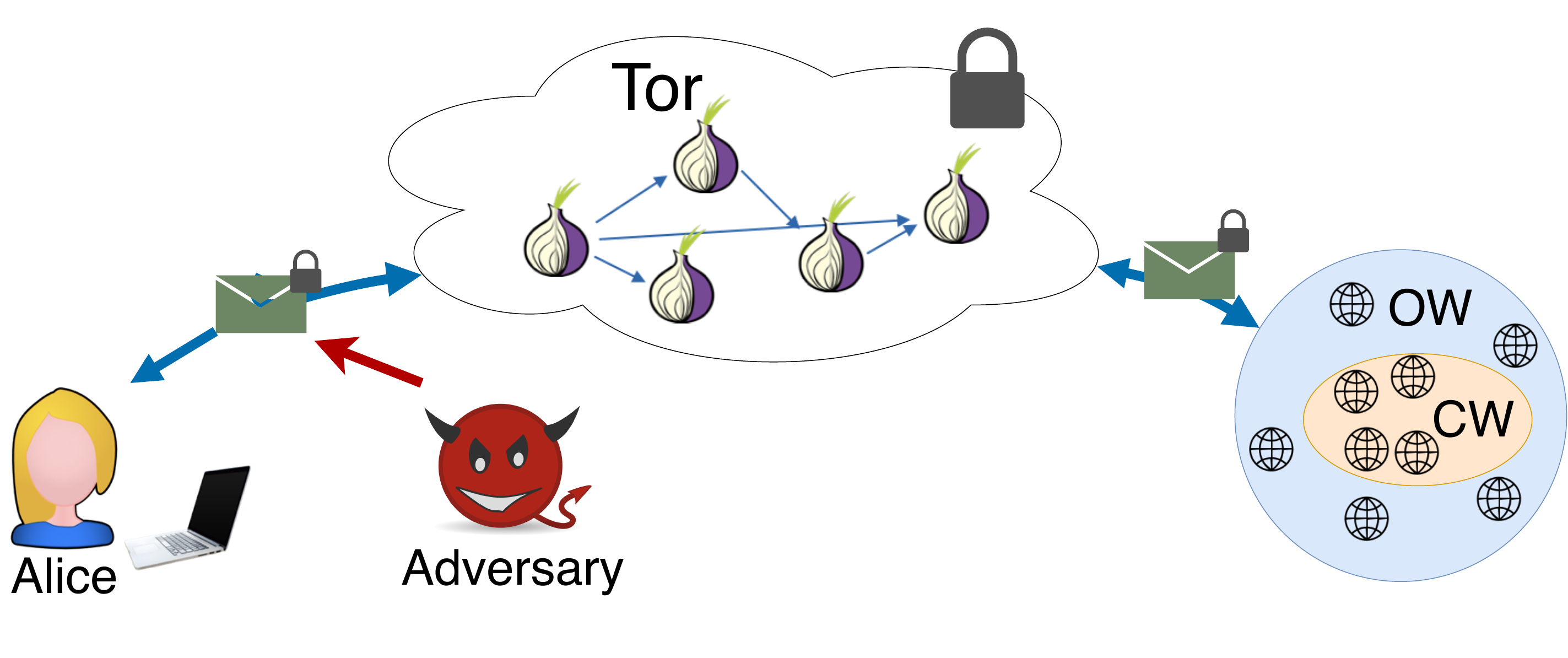}
        \vspace{-0.6cm}
        \caption{Threat model for website fingerprinting. Websites within the orange region are deemed monitored and those within the blue region correspond to unmonitored websites.}
        \vspace{-0.6cm}
        \label{fig:wf_attack_diagram}
\end{figure}

\mypara{Closed world vs open world.} The \textit{closed world} setting is considered to be the ideal scenario for an adversary. In this setting, the adversary assumes a set of websites $W = \{w_1, w_2, ... , w_n\}$ as the only websites a user is allowed to visit. The adversary then learns a model which, given any trace $t$ predicts the corresponding website $\hat{w_t} \in W$.
In contrast, the \textit{open world} setting is more realistic since an adversary assumes that a user can visit any existing website on the Internet. Since it is impossible to obtain fingerprints for \textit{all} Internet pages, the goal of the adversary is then to decide whether a client has visited a particular website amongst a set of monitored websites $W' = \{w'_1, w'_2, ... , w'_n\}$ and if so, which one.

In this paper, we choose to focus on the analysis of attacks conducted in the closed world setting. The rationale behind this choice sits on our intention to assess the security provided by a defense in the most advantageous setting for an adversary. Indeed, the closed-world setting offers the adversary a limited or complete absence of uncertainty about what websites can be potentially visited by a targeted Tor user. By focusing on this setting, DeepSE-WF analyses the most favorable scenario for an attacker, and therefore provides a lower bound for the security of any defense mechanism.

Despite the above, it is generally understood that the ability of an adversary to perform an attack in the closed world setting is unrealistic in most
practical cases~\cite{WFcritics,onlineWF}. Section~\ref{sec:conclusions} points out potential limitations of our closed world-based security estimation methodology, taking into account that we do not currently provide security estimations for defenses in the open world setting.

\section{Related Work}

This section starts by describing a range of influential WF attacks (Section~\ref{subsec:wf_attacks}) and defenses (Section~\ref{subsec:wf_defenses}). Then, we detail existing directions on the security evaluation of WF defenses (Section~\ref{subsec:wf_estimation}).

\subsection{Website Fingerprinting Attacks}
\label{subsec:wf_attacks}

One of the most important aspects of a successful website fingerprinting attack lies in the choice of the features used to train ML classifiers~\cite{AWF,wfes}. The next paragraphs present a brief description of the most influential website fingerprinting attacks on Tor developed in the last few years. Our exposition is organized according to the kind of features considered by each attack.

\mypara{Attacks using manually crafted features.} Multiple WF attacks have focused on the manual extraction of features to feed machine learning classifiers and perform predictions on user-visited websites over encrypted Tor connections~\cite{WebsiteFingerprinting,onionAnonymizationWF,touching, effectiveAttacks}. 
Among these, three attacks earned particular relevance by significantly outperforming earlier WF attempts, achieving over 90\% accuracy when classifying websites in the closed-world setting. Specifically, the k-NN~\cite{knnAttack} attack leverages the k-Nearest Neighbors classifier and a feature set with over 3\,000 features including the total transmission bandwidth and elapsed time, the number of incoming and outgoing packets, or packet ordering and bursts. The CUMUL~\cite{cumul} attack makes use of a Support Vector Machine and 104 hand-crafted features, including the number of incoming and outgoing packets and total bandwidth used in each direction. Lastly, the k-FP~\cite{kfingerprinting} attack introduces a combination of features used in previous attacks with novel traffic characteristics, leading to a systematic analysis of 150 features. The classifier works by building a fingerprint for each website using a modification of the Random Forest algorithm and then uses a k-Nearest Neighbors classifier to predict website accesses.

\mypara{Attacks using automated feature extraction.} 
Departing from the manual labour to extract features, recent research in WF attacks has shown that deep learning approaches can be successfully used to automate the feature extraction process~\cite{sdae,AWF}. 
Such attacks typically require an adversary to collect a larger amount of website traces and to find an efficient representation of these traces to train the deep learning classifier and perform a successful attack.

Constructing over the work of Abe and Goto~\cite{sdae}, Rimmer et al.~\cite{AWF} collected the largest dataset of Tor website access traces to date. They explored the performance of WF attacks when using a range of DNN-based models, including stacked-denoising autoencoders, convolutional and long short-term memory networks. Their attack, named Automated Website Fingerprinting (AWF)~\cite{AWF}, makes use of a trace packets' direction to automatically extract features.
Similarly, Oh et al.~\cite{p-fp} have also studied the ability of deep learning classifiers to launch accurate WF attacks. The Deep Fingerprinting (DF) attack proposed by Sirinam et al.~\cite{df} uses the same directional trace representation used in AWF but exploits an advanced CNN architecture that outperforms earlier deep learning attacks.  
Since then, the Tik-Tok~\cite{Tik-Tok} and Var-CNN~\cite{varcnn} attacks improved over DF by also including packet timing information. 

Recently, a related research thrust has strived to improve the success of WF when small amounts of training data are available to an adversary. Examples of such endeavors are Var-CNN~\cite{varcnn}, Triplet Fingerprinting~\cite{triplet}, GANDaLF~\cite{gandalf}, or Adaptive Fingerprinting~\cite{adaptiveWF}.

\subsection{Website Fingerprinting Defenses}
\label{subsec:wf_defenses}

Website fingerprinting defenses aim to thwart the ability of an adversary to successfully launch WF attacks by obfuscating the real characteristics of a website access trace, either by injecting dummy packets in the network, or by delaying packets according to some obfuscation scheme. 
Next, we deliver an overview of the space of existing WF defenses and their security/overhead trade-offs.

\mypara{Constant-rate padding.} Defenses like BuFLO~\cite{touching}, CS-BuFLO~\cite{csbuflo}, and Tamaraw~\cite{tamaraw} hide timing patterns and packet transmission burst behavior by leveraging different strategies that rely on the transmission of packets at fixed-rates. In addition, some of these defenses~\cite{csbuflo,tamaraw} obfuscate the size of websites being transmitted by grouping websites in sets of websites with similar sizes and padding the sites within a set to a common size. Despite their success in thwarting WF attacks, these defenses incur in large bandwidth and latency overheads that preclude their wide adoption in Tor. 
DynaFlow~\cite{dynaflow} aims to provide similar security guarantees as the above defenses, but with lower overheads. The recently proposed RegulaTor~\cite{regulator} regularizes the size of packet sequences sent over a short period of time to mask potentially revealing features.

\mypara{Supersequence.} Another class of WF defenses attempts to cluster traces of different sites to create a group of anonymity sets and extracts the shortest common supersequence. Examples of such defenses include Glove~\cite{glove} and Supersequence~\cite{effectiveAttacks}. However, the generation of supersequences requires previous knowledge about the content of websites, making these defenses hard to deploy for websites that load dynamic content. Walkie-Talkie~\cite{walkie} modifies the Tor Browser to communicate in half-duplex mode. In this way, real packets can be buffered and mixed with dummy packets to create supersequences in a more efficient fashion.

\mypara{Adaptive and randomized padding.} Adaptive padding~\cite{timingAnalysis} makes websites' packet inter-arrival timing distribution similar and indistinguishable for all traces by inserting dummy packets to mask existing time gaps between packets. WTF-PAD~\cite{wtfpad} is a lightweight adaptive padding technique tailored to Tor that exhibits a substantial overhead reduction w.r.t. earlier defenses.  FRONT~\cite{front} introduces a random number of randomly-padded dummy packets to the beginning of packet sequences to obfuscate website accesses.

\mypara{Application-layer defenses.} Defenses in this class  work at the application layer, instead of the network layer. Panchenko et al.~\cite{onionAnonymizationWF} introduced a browser plug-in that loads random websites to obfuscate a given site's traffic pattern. HTTPOS~\cite{HTTPOS} manipulates HTTP requests and the behavior of TCP to change the size and timing of packets and/or web objects.
LLaMA~\cite{lammaAlpaca} acts on the client-side and randomly delays outgoing HTTP requests while introducing dummy HTTP requests. In turn, ALPaCA~\cite{lammaAlpaca} is a server-side defense that inserts dummy web objects (or pads existing ones) to change the size of different websites to a common size.

\mypara{Traffic splitting.} This class of defenses aims to prevent an adversary from inspecting all traffic exchanged by clients, by making use of multihoming technologies to send traffic through multiple networks. Following this idea, HyWF~\cite{multihoming} splits Tor traffic towards a given Tor bridge among multiple networks. In turn, TrafficSliver~\cite{trafficSliver} splits traffic over multiple entry Tor nodes while distorting repeatable traffic patterns by distributing HTTP requests' fragments over the different paths.

\mypara{Trace merging.} These defenses rely on the difficulty to separate consecutive traces of website loads. GLUE~\cite{front} adds dummy packets that obscure the fact that two websites are loaded separately, gluing their traces together. This forces a WF adversary to correctly separate website loads before fingerprinting a website, a problem that remains  the  focus  of   research  on  multi-tab WF~\cite{hmWF,multitab}.

\mypara{Adversarial traces.} A recent class of defenses strives to thwart the success of the latest WF attacks based on deep learning techniques. Mockingbird~\cite{mockingbird} generates traces that resist WF attacks against an adversary that is assumed to be able to train a classifier in previously defended traces. Dolos~\cite{patchBasedDefenses} disrupts WF deep learning classifiers by computing input-agnostic adversarial patches that guide the injection of dummy packets into traffic traces. Nasr et al. developed BLANKET~\cite{blindPerturbations}, a technique that can defeat deep learning WF attacks by blindly perturbing the features of live connections.

\mypara{Learning-based trace generation:} Another class of defenses has started to explore the use of generative adversarial networks (GANs) to mimic realistic traffic patterns of different webpages. A prominent example from this line of work is Surakav~\cite{surakav}.

\subsection{Security Estimation of WF Defenses}
\label{subsec:wf_estimation}

So far, three main methods have been proposed in the literature to prove lower-bounds for the error of WF adversaries. We outline the three methods in terms of their chronological appearances and summarize their features, and metrics, in Table~\ref{table:estimator_features}.

\mypara{Comparative mathematical framework.} Cai et al.~\cite{tamaraw} use a comparative method for evaluating defenses against an \textit{ideal} WF adversary. 
They estimate the lower bound for the error of WF adversaries as the number of websites that produce the same network trace, and that could thus lead the adversary to erroneously classify the access to a website, irrespective of the chosen attack. To understand whether and by how much defenses are successful at mitigating WF attacks, Cai et al. first transform a website class \textit{w} into another class \textit{w'} while differing by a single feature category (e.g., packet timing). Then, they ascertain whether a defense is successful in hiding a particular feature if, after applying the defense, there is no discernible difference between \textit{w} and \textit{w'}.
Even though this method can be used to evaluate both deterministic and probabilistic defenses~\cite{walkie}, it is highly sensitive to noise in the communication (e.g., jitter), which can lead to similar traffic being misclassified~\cite{wfes}. In addition, the information leakage between features is not quantified~\cite{wefde}.

\mypara{Bayes error rate.} Cherubin~\cite{wfes} suggested to use a black-box and feature-dependent method to derive the security bounds of WF defenses by estimating the smallest achievable error, i.e., the Bayes error rate (BER), incurred by \textit{any} WF adversary. The notion of lowest possible error generalizes the framework proposed by Cai et al. \cite{tamaraw} in that the lowest error is naturally achieved if the features of all website are indistinguishable. Estimating the BER or its bounds using finite datasets is an extensively researched problem in the field of machine learning~\cite{Cover1967-zg,Fukunaga1975-zx,Buturovic1992-ij,Devijver1985-gs,Fukunaga1987-lw,Pham-Gia2007-xa,sekeh2020learning, snoopy}. 
Inspired by Cover and Hart~\cite{Cover1967-zg}, Cherubin reduces the WF problem to a classification task and leverages the error of the Nearest Neighbor classifier as a proxy to estimate the lower bound for the error of any potential classifier used on predefined features.
While this method does not depend on any specific learned classifier, assessments of the security of a given WF defense depend on the identification of a set of manually-crafted (and, ideally, optimal) feature sets. Unfortunately, the transformation of raw network traces into manually-crafted features may i) ignore traffic characteristics that provide useful information for WF attacks~\cite{df} and thus implicitly increase the BER in this transformed feature space, and ii) impose significant impacts on the convergence rates of a Nearest Neighbor classifier~\cite{knnConvergence}, jeopardizing the validity of the estimated BER lower bounds. We elaborate on this aspect in Section~\ref{subsec:reasoning} and validate it on Section~\ref{sec:compare_sota}.

\begin{table}[t]
\centering
\caption{Features and security metrics used by WF defenses' security estimators. All estimators take advantage of trace representations based on timing and direction of packets.}
\vspace{-0.4cm}
\resizebox{0.9\linewidth}{!}{%
    \begin{tabular}{lll}
        \toprule
        \textbf{Estimator}                & \textbf{Features}                       & \textbf{Metric}  \\ \midrule
        Ideal Adversary~\cite{tamaraw} & Packet Sequence  & Accuracy \\
        WFES~\cite{wfes}     & Manually crafted   & Bayes Error Rate (BER)     \\
        WeFDE~\cite{wefde}     &  Manually crafted  & Mutual Information (MI)      \\\midrule
        \textbf{DeepSE-WF (this work)}     &  Learned DL &  BER \& MI      \\\bottomrule
    \end{tabular}
}
    \label{table:estimator_features}
    \vspace{-0.3cm}
\end{table}

\mypara{Information leakage.} Li et al. proposed WeFDE~\cite{wefde}, a methodology to measure the amount of information leaked by a website fingerprint through the quantification of mutual information\footnote{The terms \textit{information leakage} and \textit{mutual information} are used interchangeably in Li et al.'s work~\cite{wefde} and in this paper.} (MI). The core idea of WeFDE is to use adaptive kernel density estimation to model the probability density function of a feature or a category of features. The estimated distributions are used to calculate the joint MI of multiple features. In order to evaluate the MI, WeFDE uses a list of 3043 features out of 14 categories, removes redundant features and groups them into clusters. Finally, it estimates the joint distribution of the clusters to get the final estimation.
Yet, and similarly to the work by Cherubin~\cite{wfes}, WeFDE relies on manually-crafted sets of features which may miss relevant traffic characteristics for informing potentially successful WF attacks.

Finally, well-established relations in terms of tight mathematical bounds between the BER and information-theoretical quantities, such as MI~\cite{fano1961transmission, kovalevsky1968problem, hellman1970probability,sason2017arimoto}, suggest that for certain regimes (e.g., perfectly secure or insecure defenses), both quantities are equally powerful to assess the security guarantees. We further discuss their relation in Section~\ref{subsec:ber_vs_mi}.
Next, we describe a set of transformations necessary to obtain features from raw network traffic, and discuss how WF attacks bypass WF defenses by relying on features that are disregarded by those defenses.

\section{Security Pitfalls due to Traffic Representation Mismatches}

In this section, we start by assembling existing work on WF attacks, analyzing the trend on how these attacks transform raw traffic data into traffic representations that are more amenable to train classifiers. Then, through the help of concrete examples found in the literature (as well as pointers to the results of our own experiments), we describe how the use of contrasting representations in attacks and defenses may lead to complications in the correct evaluation and estimation of WF defenses' security, making a case for the development of improved WF security estimation methods.

\subsection{Transformations over Network Traffic}
\label{sec:traffic_rep}

Ideally, a WF adversary would leverage all the information contained in the network traffic observations it has collected to launch successful WF attacks. In such a setting, the adversary would use a \textit{raw representation} of website loads or list every possible feature to train an effective classifier. This would however result in possibly infinite feature combinations that cannot be feasibly enumerated. 

Apart from this issue, and even if calculating such feature listing was a tractable task, prolific literature on ML~\cite{dimReduction} (and WF~\cite{wfFeatureSelection,wefde} in particular) has shown that not all features are useful for classification. In particular, and since the size of WF datasets are finite, having a large number of irrelevant and redundant features may increase the chances for overfitting and for being afflicted by  the curse of dimensionality~\cite{wfFeatureSelection}.
To tackle this issue, considerable effort has been put in place to \textit{transform} the representation of raw traffic into \textit{trace} and \textit{feature} representations that can increase the ``signal-to-noise ratio'' in the WF domain~\cite{wfFeatureSelection,wefde}.

\begin{figure}[t]
    \centering        \includegraphics[trim=0 0.8cm 0 0,clip,width=0.95\columnwidth]{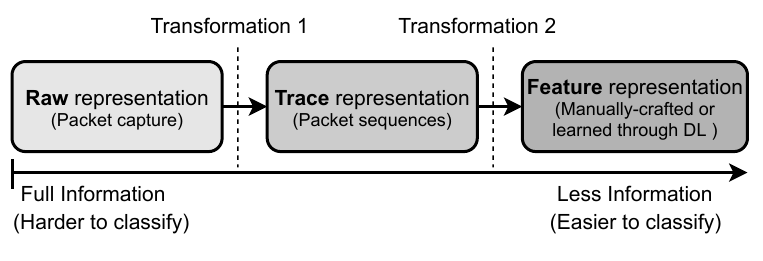}
        \vspace{-0.4cm}
        \caption{Transformations of traffic representations. Each transformation aims to summarize a traffic representation to improve the efficiency and effectiveness of WF classifiers.}
        \vspace{-0.5cm}
        \label{fig:transformations}
\end{figure}

\mypara{Transformations.} Figure~\ref{fig:transformations} depicts two usual sequential transformation steps used to process the raw representation of website loads (e.g., the contents of a \texttt{.pcap} file) into more useful representations for classification. For instance, the first transformation converts this raw representation into a simpler \textit{trace representation}, while the second transformation recasts traffic traces as a \textit{feature representation}. It should be noted, however, that each transformation incurs some information loss, e.g., losing TCP header information when obtaining a trace representation, or losing fine-grained packet timing information when obtaining a given feature representation. Nevertheless, these produce feature representations that can fuel highly effective WF classifiers.
Next, we describe the most relevant trace and feature representations derived from said transformations.

\noindent\textbf{Trace representations.} These representations are obtained from the simplification of raw traffic into packet sequences.

\begin{compactitem}
    \item \textit{Directional representation.} Represents a website trace as a sequence of 1s and -1s, resp., for outgoing/incoming packets.
    
    \item \textit{Timing representation.} Represents a website trace as a sequence of positive or negative timestamps starting from 0, depending on whether a packet is outgoing or incoming.
\end{compactitem}

\noindent\textbf{Feature representations.} These representations are obtained from the simplification of trace representations through manual feature engineering or automated feature extraction through deep learning.

\begin{compactitem}
\item \noindent\textit{Manually-crafted features.} These kind of features are obtained by extracting summary statistics from website traces, and can be split into two categories. The first category is composed of \textit{packet statistics}, where examples include the \# of incoming/outgoing packets, packet frequency, ordering, or inter-packet timing percentiles of packets in a connection.
The second category is composed of \textit{bursts statistics}, where a burst is understood as a contiguous sequence of packets sent in the same direction. Examples of features include burst size, burst duration, or the \# of incoming/outgoing bursts.

\item\noindent\textit{Learned latent features.} These features are obtained by training a DNN to project traces' representations into a latent feature space of lower dimensionality. This process enables the complete ingestion of directional/timing traffic representations, accomplishing dimensionality reduction while retaining relevant information about the original traces.
\end{compactitem}

\vspace{2pt}
Next, we expose how the inconsistent use of trace and feature representations of network traffic prevents the proper judgement of the security of WF defenses.

\begin{table}[t]
\caption{Overview of the features and traffic representations used by prominent WF attacks.}
\vspace{-0.4cm}
\resizebox{\columnwidth}{!}{%
    \begin{tabular}{p{0.18\textwidth} p{0.24\textwidth} l}
        \toprule
        \textbf{Attack}            & \textbf{Features}      & \textbf{Trace Representation}                                                                     \\ \midrule
        (2014) kNN~\cite{knnAttack}  & Manually crafted  & Time, Direction     \\
        (2016) CUMUL~\cite{cumul}    & Manually crafted  & Time, Direction     \\
        (2016) kFP~\cite{kfingerprinting}  & Manually crafted  & Time, Direction \\\midrule
        (2017) AWF~\cite{AWF}  & Learned DL     & Direction \\
        (2018) DF~\cite{df}   & Learned DL      & Direction  \\
        (2019) TF~\cite{triplet}  & Learned DL   & Direction\\
        (2021) Adaptive WF~\cite{adaptiveWF}  & Learned DL    & Direction  \\
        (2021) GANDaLF~\cite{gandalf}  & Learned DL (GAN)   & Time, Direction \\\midrule
        (2018) Var-CNN~\cite{varcnn}  & Learned DL + Manually crafted & Time, Direction  \\
        (2020) Tik-Tok~\cite{Tik-Tok}  & Learned DL   & Time, Direction \\
        \bottomrule
    \end{tabular}
}
    \label{table:attack_features}
    \vspace{-0.5cm}
\end{table}

\subsection{Flawed Security Estimations due to Traffic Representation Inconsistencies}

One of the most noticeable trends in the development of new WF attacks is the importance attributed to the analysis of alternative traffic representations and to the development of comprehensive feature extraction processes -- now believed to be \textit{the most critical step} in devising successful WF attacks~\cite{wfes}.
This trend is better observed in Table~\ref{table:attack_features}, which depicts a breakdown of the features and trace representations used by some of the most prominent attacks in the WF literature. The table helps us identify three main categories of attacks, depending on the nature of features, namely whether they are i) manually crafted, ii) learned through deep learning, or iii) a mix of both. Further, Table~\ref{table:attack_features} highlights that attacks based on manually crafted features tend to rely on both types of trace representations (time and direction), whereas DL-based attacks sometimes tend to rely on direction only with recent advances incorporating the timing information again.

A closer analysis of this table results in two main observations, which reveal the inadequacy of a large fraction of existing WF defenses and security estimators to cope with the latest WF attacks:

\mypara{(1) The most recent attacks tend to use as much information as possible about website traces.}
As revealed by recent WF attacks like Tik-Tok or Var-CNN, the use of timing information allows for an increased success in performing website fingerprinting. This means that the security properties provided by defenses that exclusively attempt to hide directional patterns, e.g., to defend against AWF or DF, may be severely degraded when facing a WF adversary that makes use of directional and timing trace representations, and the additional information contained therein.

As a concrete example, consider Walkie-Talkie~\cite{walkie}, a defense that relies on padding two websites such that their directional trace representation looks the same. Thus, an attack based on directional information would achieve an accuracy of at most 50\% (this was indeed approximated by the DF attack, which obtained 49.7\%). However, the Tik-Tok attack proposed by Rahman et al.~\cite{Tik-Tok} has shown that including packet timing information can increase the success of fingerprinting Walkie-Talkie defended traces up to 97\%.

The above re-emphasizes the current \textit{status quo} in learning-based traffic analysis; attacks should use as much information as possible to increase their effectiveness over WF defenses, particularly those that are able to modify both directional and timing aspects~\cite{varcnn,Tik-Tok}.

\begin{figure*}[t]
    \centering
    \includegraphics[width=0.9\textwidth]{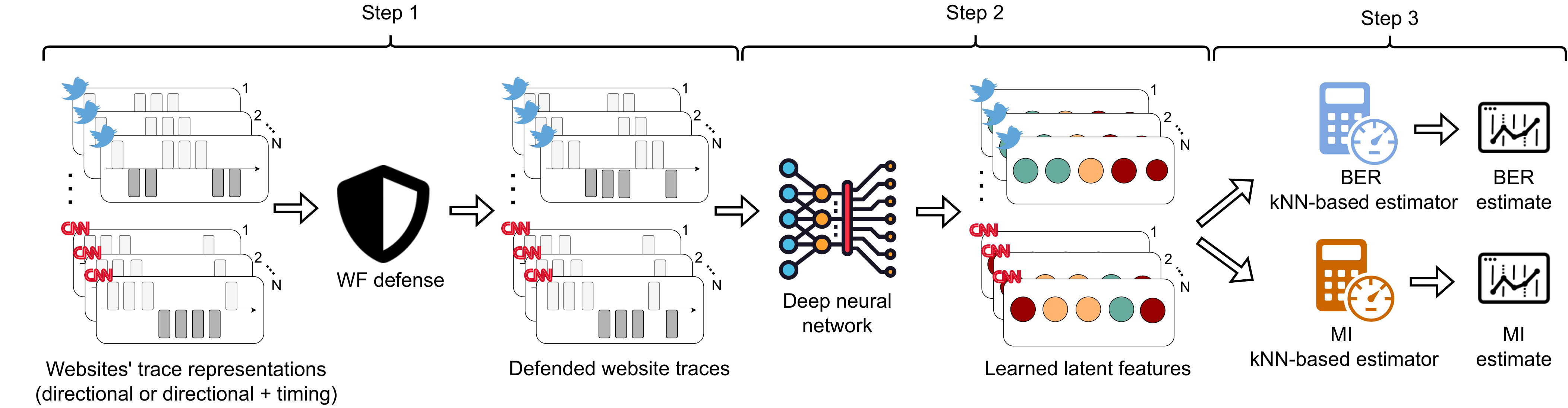}
    \vspace{-0.3cm}
    \caption{DeepSE-WF pipeline for the unified security estimation for WF defenses. We start by applying a specific WF defense on the rawest compatible trace representation (directional, or directional and timing). We then train a DNN on the defended traces and take the minimal/maximal estimation over kNN-based estimations of the Bayes error/MI, respectively.}
    \label{fig:unified_estimation}
    \vspace{-0.4cm}
\end{figure*}

\mypara{(2) WF attacks no longer make an extensive use of manually-crafted features.}
As shown in Table 2, WF attacks departed from the exclusive use of manually crafted features -- which may unwillingly fail to include important features about website traces -- and currently leverage feature extraction mechanisms based on DNNs. This enables the generation of latent feature vectors representing complex features that support effective learning, by distilling the large amounts of information contained in trace representations.

In turn, this means that \textit{existing security estimation methods can no longer be trusted to perform accurate estimations}. As a matter of fact, security estimators like WFES and WeFDE (see Table~\ref{table:estimator_features}) focus solely on the analysis of manually-crafted features, whereas deep learning-based attacks extract latent features that are much more complex than those obtained via manual feature engineering methods based on expert knowledge. Indeed, as shown in our evaluation (Section~\ref{sec:compare_sota}), the classification error achieved by state-of-the-art WF attacks (e.g., Tik-Tok) is substantially smaller than the estimates provided by WFES, for all defenses under test that are not based in the creation of constant-rate flows.

We conjecture that this discrepancy between attacker accuracy and security estimation is mainly given by the aforementioned mismatch in feature representations, and indirectly by the finite, fixed size datasets used to provide security estimates.

Next, we introduce a new security estimation method for WF defenses which considers latent feature spaces and is able to bridge the gap between earlier estimation methods and current WF attacks.

\section{Improved Security Estimations for Website Fingerprinting Defenses}
\label{sec:improved_estimations}

This section introduces DeepSE-WF, a novel framework that produces tight estimations of the security of WF defenses. Our framework explicitly targets the security evaluation of defenses in the simplified closed-world WF scenario (Section~\ref{sec:wf_background}). This scenario allows an adversary to launch a WF attack in controlled settings, thus making it ideal to evaluate the security of defenses~\cite{wfes,WFcritics}.
Next, we provide an overview of our estimation methodology, describe the architecture of DeepSE-WF, and provide the theoretical reasoning guiding our approach.

\subsection{Estimation Methodology}
\label{sec:estimation_methodology}

DeepSE-WF is a new WF security estimation method relying on specialized BER and MI estimators that leverage learned latent feature spaces generated by deep learning approaches. In contrast to existing WF security estimation frameworks that make use of manually-crafted features, the use of latent feature spaces allows DeepSE-WF to level the playing ground w.r.t. to the features used by DNN-based WF attacks (e.g.,~\cite{df,Tik-Tok}).

\mypara{Estimating the Bayes error.} Bayes error estimators deployed over different feature transformations have been extensively studied by Renggli et. al.~\cite{renggli2021evaluating}. A key finding states that a very simple estimator given by Cover and Hart~\cite{Cover1967-zg} consistently outperforms all other newer estimators when run over pre-trained transformations instead of the raw features. The estimator uses the nearest neighbor (NN) error to estimate the BER. If there is a lack of  \textit{good} pre-trained transformations available,
one can train such a representation to increase the performance of the BER estimator; the training process should be performed on a different dataset than the one used to actually estimate the BER to avoid overfitting. An intuitive way to train such a representation would be to train a representation which at the same time minimizes the kNN error rate. As the calculation of this value is not differentiable, one has to either approximate the kNN error~\cite{goldberger2004neighbourhood}, or minimize another differentiable classifier's error. We chose the latter and thus minimize the cross entropy loss of a linear classifier, supported by the fact that a lower linear classifier error translates to better kNN performance~\cite{knnConvergence}.

\mypara{Estimating mutual information.} Estimating the MI is a vibrant area of research in the field of ML and information theory. One prominent area of research aims to improve simple kNN-based estimators over raw features to perform well on combinations of discrete and continuous datasets~\cite{gao2017estimating}. As these approaches often fail in high-dimensional spaces, another line of work pioneered by MINE~\cite{mine} simultaneously learns complex representations and estimates the MI on high dimensional inputs. As we discuss in Appendix~\ref{app:mine}, it is non-trivial to train MINE on unexplored data modalities, mainly due to the hard task of choosing a good set of hyperparameters or working on limited amounts of data. Thus, DeepSE-WF relies on a (to the best of our knowledge) novel approach to estimate the MI which, inline with the BER estimation strategy, consists of deploying a simple kNN-based estimator on top of separately trained representations.

\subsection{The Architecture of DeepSE-WF}
\label{sec:deepse_architecture}

Figure~\ref{fig:unified_estimation} depicts DeepSE-WF's security estimation pipeline. In the first step of the pipeline, website traces are first transformed, i.e., defended, with the help of a WF defense. In this paper, we generate defended traces by using defense simulators based on pre-recorded traces, with further details on the use of simulated defended traces in Section~\ref{sec:evaluation_setup}. Despite
this choice, nothing precludes DeepSE-WF to be used to perform security estimations based on real defended
traces obtained with actual defense implementations, i.e., DeepSE-WF is agnostic to the way defended
traces are generated.

Then, in the second step of the pipeline, DNNs are trained to generate feature representations of defended packet sequences in a latent feature space on many different trace representations. Finally, the resulting learned latent features are used as input to special-purpose kNN-based estimators, which estimate the BER and the MI. Here, we take the minimum and maximum over trace representations, respectively, for estimating the BER and MI of the defended traces.
We now shed light over each step of our estimation pipeline:

\mypara{Step 1 - Collect network observations and apply a WF defense mechanism.} First, our pipeline applies a given WF defense on trace representations of network traffic (i.e., both directional only, and direction+timing). Then, it splits the resulting data $X$ into two disjoint sets $\mathcal{T}$ (training set) and $\mathcal{E}$ (testing set). This step ensures that our security estimations do not overfit to any finite-sample dataset, a natural problem of deep learning, or machine learning in general. The sizes of the splits are given in Section~\ref{sec:evaluation_setup}.

\mypara{Step 2 - Train and extract deep learned features from multiple trace representations.} The second step of our pipeline learns feature representations through the training of a DNN using $\mathcal{T}$. 
To make DeepSE-WF compatible with both directional and directional + timing representations, we train a DNN for each of the two traffic representations. We then extract the last-layer features for all trained DNNs.
In practice, we use a DNN that ingests packets' direction and timing (similar to Tik-Tok) to evaluate defenses that consider both traces’ representations (e.g., Tamaraw), and a DNN that solely considers directional traces (similar to DF) to evaluate defenses that only consider these traces (e.g., Walkie-Talkie).

\mypara{Step 3 - Perform kNN-based security estimations based on the latent feature spaces.} The third step of our pipeline performs and delivers BER and MI security estimates for WF defenses. To compute a minimum BER estimate, DeepSE-WF relies on formula:

\begin{equation}
\label{eq:EstimateBer}
\min_{f} \widehat{(R_{f(X)})}_{n, 1} = \min_{f} \left( \frac{(R_{f(X)})_{n, 1}}{1+\sqrt{1-\frac{C (R_{f(X)})_{n, 1}}{C-1}}} \right),
\end{equation}
where $(R_{f(X)})_{n, 1}$ is the kNN error with $k=1$ attained after transforming the features $X$ with function $f$, and $C$ is the number of classes. $f$ is either the trained deep feature representation on top of directional traces, or the one trained using direction and timing.

Correspondingly, DeepSE-WF delivers the maximum MI estimate, with formula:
\begin{align}\label{eq:EstimateMI}
\begin{split}
\max_{f} \hat{I}(f(X);Y) = \max_{f} &\left( \psi(N) - \left< \psi(N_x) \right> \right. \\ &+ \left. \psi(k) - \left< \psi(m_f) \right>\right),
\end{split}
\end{align}
with the same feature transformations as in Eq.~\eqref{eq:EstimateBer} and $\psi$ denoting the digamma function; $N$ represents the total number of samples, $N_x$ is the number of samples per class averaged over all classes; $k$ is a hyperparameter (normally chosen to be small, 5 in our case), and $m$ captures the average number of samples (class-independent) in the radius defined by the $k$ nearest samples of the same class for every data point. The full details about the kNN-based estimator in Eq.~\eqref{eq:EstimateMI} can be found in the original paper by Ross~\cite{ross2014mutual} (c.f., Eq.~\eqref{eq:EstimateMI} therein). Notice that the only transformation-dependent parameter is $m$ and the estimator is known to be asymptotically unbiased when reporting \textit{nats} (i.e., a natural unit of information assuming the natural logarithm for the information theoretical quantity)~\cite{gao2015efficient}. We thus multiply the result of Eq.~\eqref{eq:EstimateMI} by a constant factor $\log_2(e)$ in all experiments to have \textit{bits} as a resulting unit.
We refer the reader to Appendix~\ref{app:mine} for more details on the possible use of alternative MI estimation techniques in the context of DeepSE-WF.

\subsection{Theoretical Reasoning}
\label{subsec:reasoning}

Our method is theoretically guided by three main observations, which we detail in the following paragraphs.

\mypara{(1) WF defenses rely on transformations which purposely decrease useful information.} Since WF defenses are typically designed as transformations acting solely on features, we examine the impact of such feature transformations on two quantities of interest. First, it is well known that any deterministic feature transformation can only increase the BER~\cite{knnConvergence}. Conversely, it is easy to show (we provide a proof in Appendix~\ref{app:proof_mi_transformations}) that any deterministic feature transformation can only decrease the MI between features and labels.  This is precisely what a WF defense is trying to achieve: a good defense is represented by a transformation which purposely increases the BER towards its maximum $\frac{C-1}{C}$, where $C$ are the number of classes, and simultaneously decreases the mutual information towards $0$ whilst having low computational overhead.

\mypara{(2) Estimating security guarantees on raw representations is hard.} Guided by observation (1) and the fact that assessing the potential increase of the BER, or decrease in MI induced by any transformation, or by a chain of transformations, is equally hard as estimating the BER or MI itself, it is natural that one should take the \textit{rawest} possible representation\footnote{Injective transformations are safe, as they are invertible and thus intuitively do not lead to any reduction in information~\cite{knnConvergence}.}, i.e., not performing any additional change in representation, to estimate security guarantees after having applied the WF defense. Still, as we are estimating a property of an unknown probability distribution (i.e., for an infinite amount of data) based on a finite-sample dataset, we have to consider the convergence behavior of the estimators.

We now outline this reasoning for a theoretically understood BER estimator.
Having access to infinite samples, one could use a consistent classifier (i.e., one with its error converging to the BER) like the k-Nearest-Neighbor (kNN) estimator, which is known to be strongly consistent provided $k$ diverges (i.e., grows to infinity), whereas $\frac{k}{n}$ converges to $0$ as $n \rightarrow \infty$, where $n$ is the number of samples.
Formally, the convergence of the kNN classifier error $(R_{X})_{n,k}$ is given by:
\begin{equation}\label{eqnConvRatesRaw}
\mathbb{E}_n \left[ (R_{X})_{n,k}\right] - R_{X}^*= \mathcal{O} \left( \frac{1}{\sqrt{k}}\right) + \mathcal{O}\left( L\left( \frac{k}{n}\right)^{1/D} \right),
\end{equation}
where $D$ is the raw feature dimension, $L$ is some distribution dependent constant, and $R_{X}^*$ is the BER~\cite{Gyorfi2002}.

In other words, the kNN algorithm converges to the BER with increasing number of samples, as long as $k$ is also increased, but at a slower rate than the number of samples.
Unfortunately, we have neither access to infinite number of samples, nor can we set $k$ to be infinitely large. One alternative approach is therefore to estimate bounds on the BER rather than the BER itself.

The BER estimator proposed by Cover and Hart~\cite{Cover1967-zg}, uses the NN accuracy (i.e., the kNN accuracy for $k=1$) as an upper bound of the BER\footnote{In fact any expected classifier accuracy trained on any number of samples is a valid upper bound of the BER.}, and a scaled version of the NN accuracy $\widehat{(R_X)}_{\infty, 1}$ to estimate the BER lower bound. Formally, under mild assumptions we have:
\vspace{-0.1cm}
\begin{equation}\label{eqnCoverHart}
\begin{small}
(R_X)_{\infty,1 } \geq  R_{X}^* \geq \frac{(R_X)_{\infty, 1}}{1+\sqrt{1-\frac{C (R_X)_{\infty, 1}}{C-1}}} = \widehat{(R_X)}_{\infty, 1}.
\end{small}
\end{equation}

Notice that this lower bound is only guaranteed to be \textit{valid} for infinite number of samples. For a fixed finite $n$, the lower bound estimate $\widehat{(R_X)}_{n, 1}$ can in fact be \textit{wrong} even in expectation (i.e., neglecting the impact of the variance). The regime in which the estimator is wrong depends on the ratio between finite-sample positive bias (i.e., $\widehat{(R_X)}_{n, 1} - \widehat{(R_X)}_{\infty, 1})$, given by the convergence of the estimator, and the tightness of the lower bound for a fixed probability distribution (i.e., $R_{X}^* - \widehat{(R_X)}_{\infty, 1}$).

Given that the finite-sample bias in all examined real-world use-cases is strictly larger than the tightness of the lower bound estimate~\cite{renggli2021evaluating}, we refer to the lower bound estimator in Eq.~\eqref{eqnCoverHart} for the remainder of the paper as a BER (and not a lower bound) estimator.
The convergence rate of the NN-based estimators is known to have an exponential dependency on the dimension of the raw feature space (c.f., Eq.~\eqref{eqnConvRatesRaw}). Still, this BER estimator has been shown to be very powerful (i.e., reducing significantly the finite-samples bias) when applied on top of (potentially pre-trained) feature transformations~\cite{renggli2021evaluating}. The convergence rate of the kNN classifier error on top of any feature transformation becomes:
\vspace{-0.1cm}
\begin{align}\label{eqnConvRatesThm}
\begin{split}
\mathbb{E}_n \left[ (R_{f(X)})_{n,k}\right] &- R_{X}^*= \\
&\mathcal{O} \left( \frac{1}{\sqrt{k}}\right) + \mathcal{O}\left( L_g \left( \frac{k}{n}\right)^{1/d} \right) + \delta_f,
\end{split}
\end{align}
where $d$ is the dimension of the transformed features, $L_g$ a property of the transformed feature space, and $\delta_f$ represents the increase of the BER~\cite{knnConvergence}. Thus, despite potentially having another small positive bias term (i.e., an increase in BER at infinity), transformations can speed up the finite-sample convergence of the classifier accuracy if $L_g$ and $d$ are much smaller than $L$ and $D$ respectively, and thus positively impact the accuracy of the BER estimator.

Given that existing consistency and convergence analyses for kNN-based MI estimators, which represent the most studied MI estimators~\cite{kozachenko1987sample, kraskov2004estimating, gao2017estimating}, deployed with finite $k$ or applied on transformed data are largely missing, and outside of the scope of this work, we conjecture that the same reasoning holds for the simple kNN-based MI estimator introduced by Ross~\cite{ross2014mutual} by substituting the terms ``BER'' and ``positive bias'' with ``MI'' and ``negative bias'' for the remaining of this section.

\mypara{(3) Better estimators can be obtained by observing different feature transformations.} Guided by the reasoning in observation (2) and the fact that the sum of finite-sample bias and transformation bias are positive and typically strictly larger than the tightness of the lower bound estimate, it is natural that even without having knowledge of the additional induced biases the NN-based estimators can only benefit from deterministic feature transformations in the finite-sample case. A natural consequence of this is that one can simply achieve a better estimate of the BER by not only inspecting a single representation, but rather try many different transformations and report the minimal achieved BER~\cite{snoopy}. Note that a theoretical counterpart for NN-based MI estimators (e.g., negative finite-sample bias) is missing. Nevertheless, our evaluation in Section~\ref{sec:evaluation} suggests that for the evaluated defenses and dataset sizes, one can take the maximal estimate kNN-based MI score over various transformations without noticing any impact of such a negative bias.

\mypara{Takeaways. }
All in all, the theoretical reasoning from before helps us motivate two major aspects of DeepSE-WF: (1) why using a BER or MI estimate for assessing the effectiveness of WF defenses is the metric of interest and better suited compared to using the accuracy of WF attacks, and (2) why kNN-based estimators can safely be deployed over multiple transformations. The latter, backed with the given convergence rates of BER estimators, can be useful for future research, where sample complexity and convergence for equally performing defenses could be compared based on synthetically generated traces with known BER.

\section{Experimental Evaluation}
\label{sec:evaluation}

This section details the experimental evaluation of DeepSE-WF on multiple axes. We start by describing our experimental setup (Section~\ref{sec:evaluation_setup}). Then, we deliver a comparison of DeepSE-WF security estimates against existing estimation methods (Section~\ref{sec:compare_sota}), analyse the convergence of our proposed estimators (Section~\ref{sec:knn_convergence_eval}), and discuss DeepSE-WF's results when using alternate DNN architectures and parameterizations when producing security estimates (Section~\ref{sec:dnn_arch_params}). Lastly, we perform additional validations on DeepSE-WF's estimates using an alternative dataset (Section~\ref{sec:ds19_dataset}).

\subsection{Experimental Setup}
\label{sec:evaluation_setup}

This section describes our assumptions and evaluation testbed.

\mypara{Assumptions.} In our evaluation, we assume that website traces are i.i.d. random variables, that is, we assume that all traces sampled from a given website follow the same arbitrary distribution. Previous work showed that violating this assumption would only deteriorate a WF adversary's success when attempting to fingerprint a website~\cite{wfes}. Moreover, we assume accesses to any monitored website (in the closed-world setting focused in our work) to be equally likely. We also assume a very powerful attacker who can perfectly separate the traces pertaining to the loading of different websites and that is able to find out which defense is under use.

\mypara{Dataset.} The bulk of our experiments use the dataset released by Rimmer et al.~\cite{AWF}. The closed-world dataset consists of over 900 unique websites where each website is visited up to 5000 times using the Tor network. We filter out corrupted traces (i.e. traces beginning with an ACK or incoming packet) and sort the packets according to their timestamp. We remove all packets which do not carry any Tor payload and extract directional and timing-related information. Each trace is truncated to a length of 5000 packets or padded with zeros if the original trace is shorter. We refer to the dataset as AWF$_{W}._{T}$, where \textit{W} refers to the number of unique websites and \textit{T} to the number of traces per website in the dataset. We perform additional validations with the DS19 dataset~\cite{front}.

\mypara{Cross-validation.} We  make use of cross-validation to produce accurate security estimations in our experiments. We split the data into five splits, where each split contains the same number of traces per website. Four splits are used to train the classifier and the fifth split is divided into two test sets, $\mathcal{E}_1$ and $\mathcal{E}_2$ of equal size and equal number of traces per website. We generate DeepSE-WF estimations using $\mathcal{E}_1$ for training the kNN classifier and testing on $\mathcal{E}_2$ and vice versa, where we average the results. Since we split the data and do not use any training data in the BER/MI estimation, we mitigate any unwanted bias in the estimation. This process is repeated using the next split for the BER/MI evaluation and the rest for training the classifier. We experience a negligible variance in our evaluation ($\leq 1\%$) and thus omit the confidence interval in our plots.

\mypara{Attacks, defenses, and estimators.} We use publicly available code to i) simulate WF attacks, ii) generate defended traces, and iii) estimate the security of WF defenses. Specifically, we use Cherubin's codebase~\cite{wfesImpl} to experiment with the Tamaraw and CS-BuFLO defenses, as well as the WFES estimator; Gong et al.'s repository~\cite{frontglueImpl} to experiment with the FRONT and WTF-PAD defenses; and Rahman et al.'s re-implementation of WeFDE~\cite{rewefdeImpl}. 
We also implemented the kNN-based BER estimator proposed by Renggli et al~\cite{snoopy}.

We analyse a selection of WF defenses that assume the ability of an adversary to inspect all of a client's traffic links (e.g., in contrast to TrafficSliver~\cite{trafficSliver}), that do not assume the protection of directional information-only (e.g., in contrast to Mockingbird~\cite{mockingbird} or Dolos~\cite{patchBasedDefenses}), and that provide a defense simulator that can be applied to existing traffic traces (e.g., differently from BLANKET~\cite{blindPerturbations}).
We use the original attacks and defense simulators with their recommended parameters, i.e., we perform no parameter tuning (Appendix~\ref{sec:hyperparams}).
We note that defense simulators aim to produce a faithful representation of how defenses should operate on real traffic. Typically, undefended traces are collected on the real Tor network, and then modified in an ``offline'' way, resulting in defended traces based on the theoretical outcome of the defense. Gong et al.~\cite{WFDefProxy} have recently compared the results obtained by the simulation and implementation of a set of WF defenses, finding that simulations correctly capture the strength of each defense against attacks.

\mypara{Laboratory testbed.} To assess the computational performance of DeepSE-WF when compared to existing estimators, we used a MacBook Pro with an M1 Pro CPU and 32 GB  RAM. To evaluate further performance enhancements when using a GPU, we used an Ubuntu 20.04 machine provisioned with 40 2.1 GHz Intel Xeon E5-262 CPU cores, an NVIDIA TITAN X GPU and 256 GB RAM.

\begin{figure*}[t]
    \centering
    \begin{subfigure}[t]{0.4\textwidth}
        \includegraphics[trim={0 0 0 2.4cm},clip,width=\textwidth]{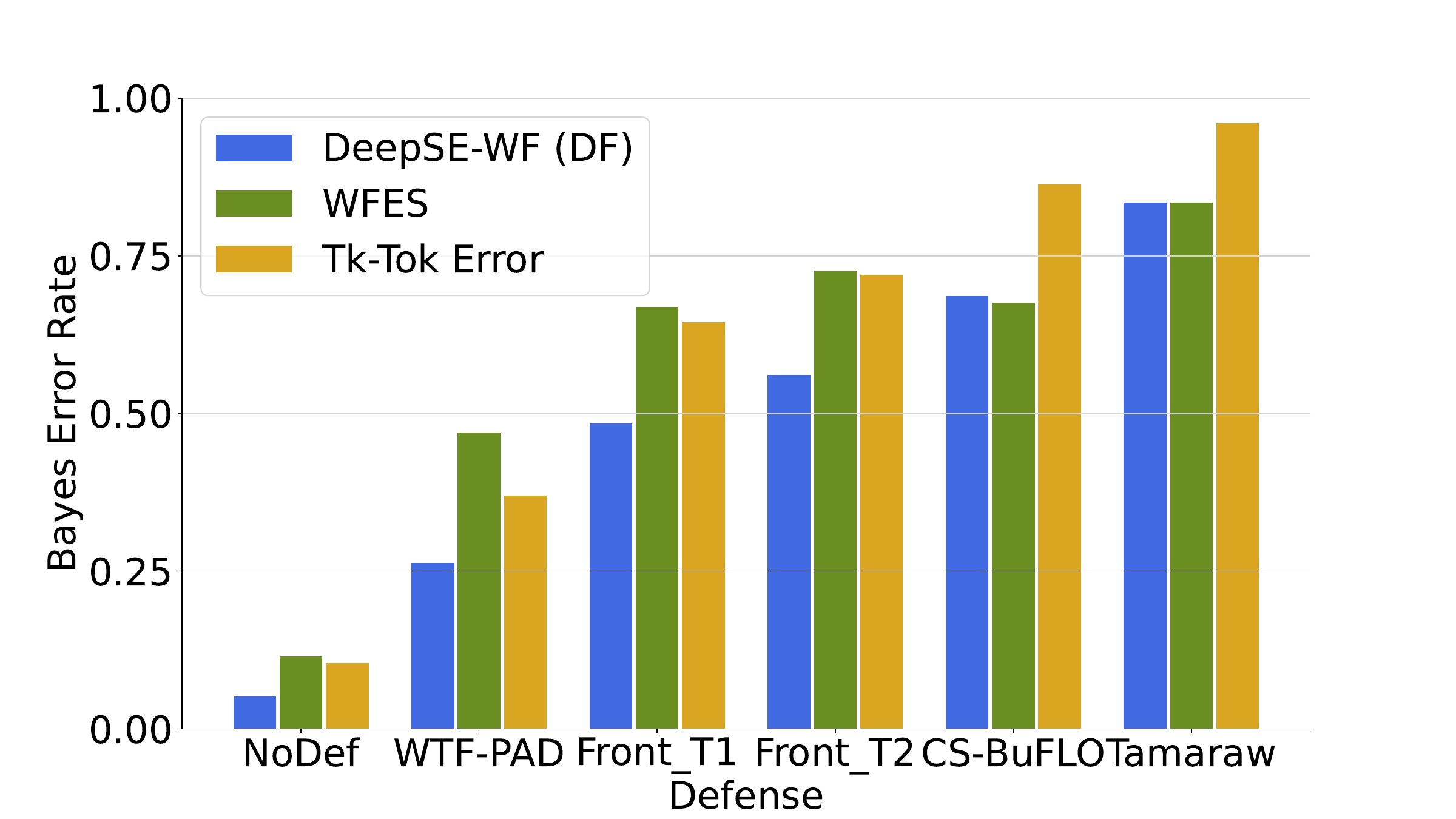}
        \vspace{-0.6cm}
        \caption{BER estimated by DeepSE-WF and WFES (AWF$_{100}._{90}$). Tik-Tok's error shown for comparison.}
        \label{fig:ber_comparison_90}
    \end{subfigure}
    \hspace{0.4cm}
    \begin{subfigure}[t]{0.4\textwidth}
        \includegraphics[trim={0 0 0 2.4cm},clip,width=\textwidth]{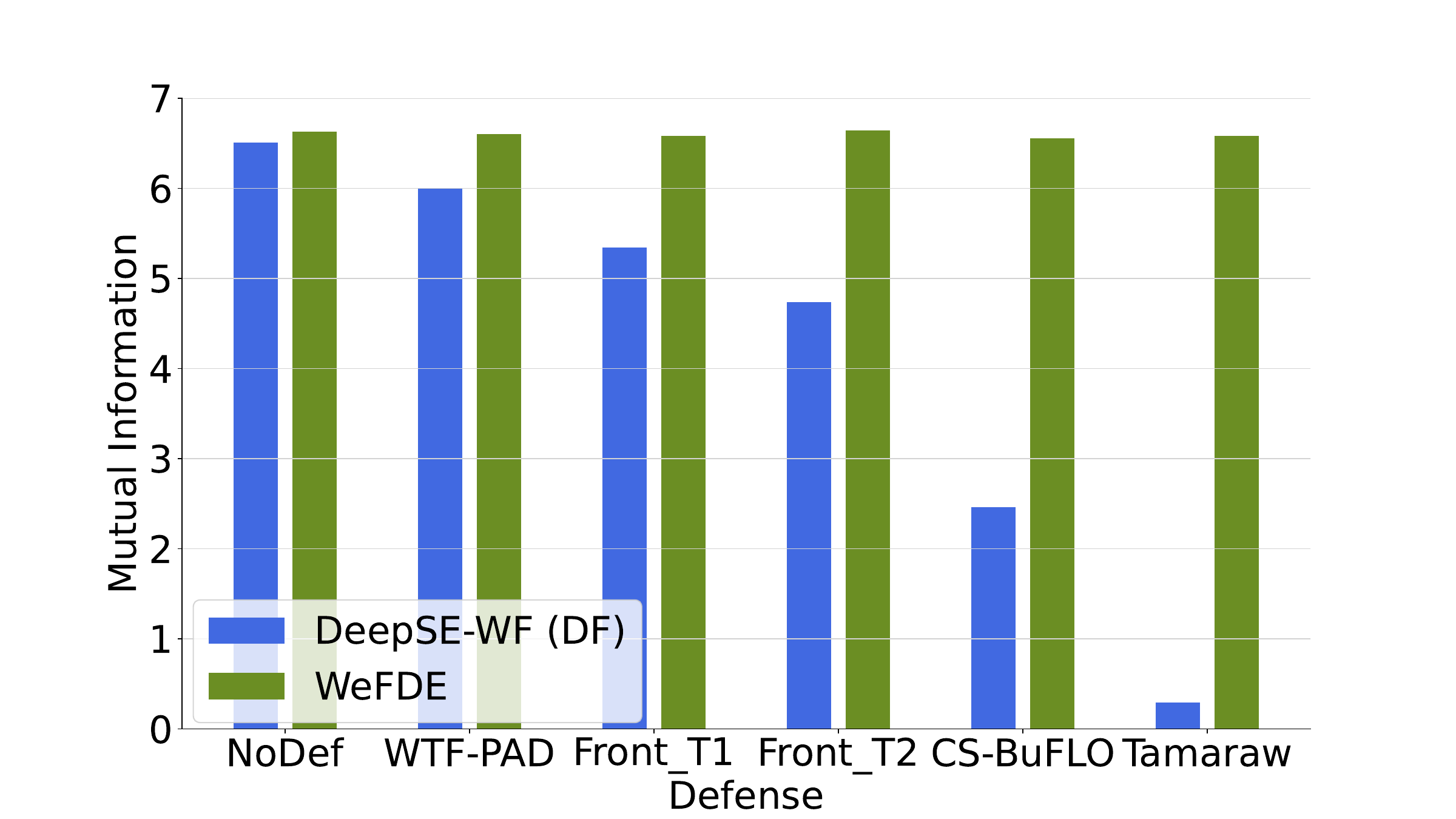}
        \vspace{-0.6cm}
        \caption{MI estimated by DeepSE-WF and WeFDE (AWF$_{100}._{500}$).}
        \label{fig:mi_comparison_500}
    \end{subfigure}
    \vspace{-0.4cm}
     \caption{BER and MI estimates obtained by DeepSE-WF and other WF security estimators. 
     }
     \label{fig:estimationBoundComp}
     \vspace{-0.2cm}
\end{figure*}

\begin{figure*}[t]
    \centering
    \begin{subfigure}[t]{0.4\textwidth}
        \includegraphics[trim={0 0 0 2.4cm},clip,width=\textwidth]{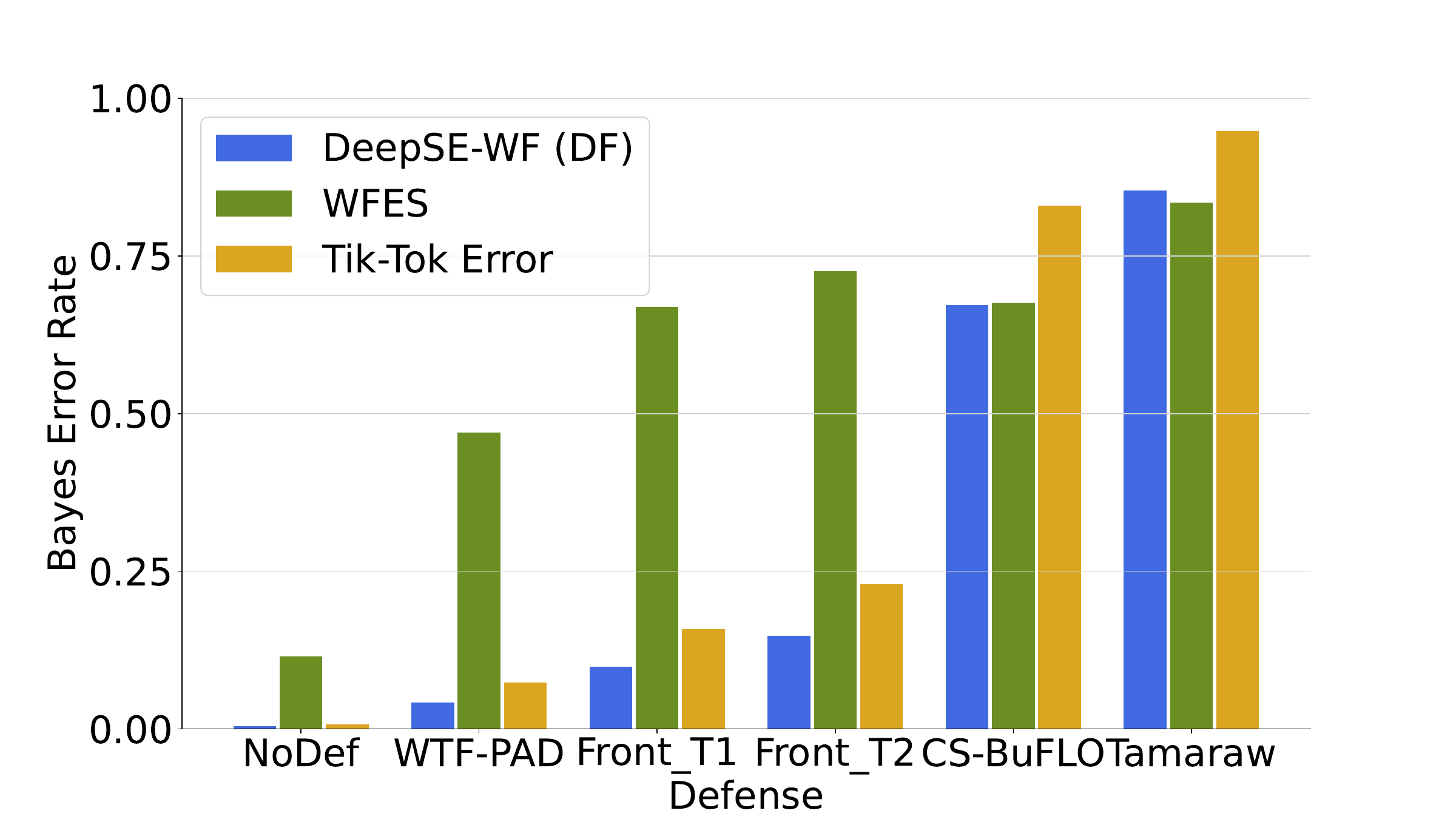}
        \vspace{-0.6cm}
        \caption{BER estimated by WFES (AWF$_{100}._{90}$). DeepSE-WF and Tik-Toks's error (AWF$_{100}._{4500}$) shown for comparison.}
        \label{fig:ber_comparison_best}
    \end{subfigure}
    \hspace{0.4cm}
    \begin{subfigure}[t]{0.4\textwidth}
        \includegraphics[trim={0 0 0 2.4cm},clip,width=\textwidth]{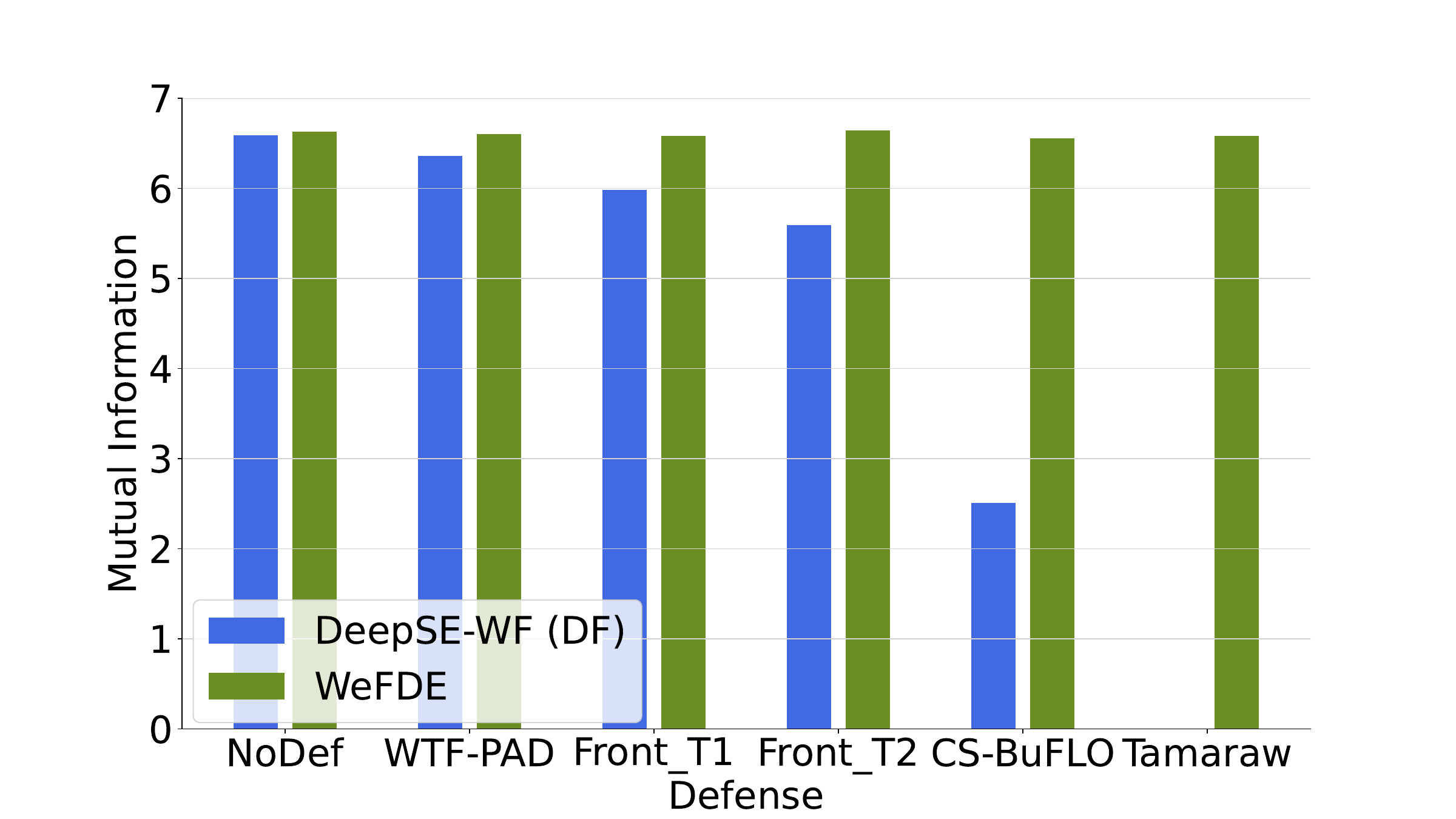}
        \vspace{-0.6cm}
        \caption{MI estimated by WeFDE (AWF$_{100}._{500}$). DeepSE-WF (AWF$_{100}._{4500}$) shown for comparison.}
    \label{fig:mi_comparison_best}
    \end{subfigure}
    \vspace{-0.4cm}
    \caption{BER and MI estimates as obtained by DeepSE-WF and other WF security estimators when additional data is available.}
    \label{fig:estimationScaleComp}
    \vspace{-0.4cm}
\end{figure*}

\subsection{Comparison to Existing Estimators}
\label{sec:compare_sota}

We relied on the DF architecture to generate BER estimates and we compared DeepSE-WF (DF) against WFES and WeFDE, the state-of-the-art BER and MI estimators, respectively, in three different dimensions: \textit{tightness of estimates}, \textit{scalability}, and \textit{computational performance}. While the original DF consumes directional traces only, we extended its architecture to ingest directional + timing traces  (corresponding to Tik-Tok's extended architecture). Next, we present our results and main findings.

\mypara{DeepSE-WF delivers tighter estimates.} Figure~\ref{fig:estimationBoundComp} depicts the comparison between DeepSE-WF's BER and MI estimations against the ones obtained by WFES and WeFDE. We leverage the AWF$_{100}._{90}$ and AWF$_{100}._{500}$ datasets, respectively, since these are the maximum dataset sizes we can process with reasonable memory and time requirements on WFES and WeFDE with our laboratory machine. For WFES, we estimate the bounds for all implemented classifiers and report the smallest one achieved (see all computed bounds in Appendix~\ref{app:extended_results}). Figure~\ref{fig:estimationBoundComp} a) reveals that, except for CS-BuFLO and Tamaraw, DeepSE-WF achieves a lower BER estimation than that provided by WFES. We also see that the Tik-Tok error is below the WFES estimates for all defenses which are not using constant rate traffic. 
Our results also show that there is a significant margin for the improvement of WF attacks since a large gap exists between DeepSE-WF and Tik-Tok error, e.g., as large as $\approx$0.18 for Front\_T1. In addition, Figure~\ref{fig:estimationBoundComp} b) reveals an interesting insight regarding WeFDE: despite its usefulness in estimating per-feature information leakage, it does a poor job at jointly estimating the MI when all features are involved (close to the maximum of $\log_2(100) \approx 6.65$ bits) regardless of whether a defense is applied. In this regard, the figure shows that the MI estimations of DeepSE-WF produce more reasonable results compared to WeFDE. For example, WeFDE estimated the information leakage of CS-BuFLO to be close to the maximum, suggesting that there exists an attack which is able to perfectly classify websites when the defense is used (in contrast to Figure~\ref{fig:ber_comparison_90}). DeepSE-WF estimates the information leakage to be only 2.46 bits, which seems to be a more reasonable estimation.

\mypara{DeepSE-WF is more scalable.} As mentioned in the previous paragraph, WFES and WeFDE exhaust our testbed machine's memory when using datasets larger than AWF$_{100}._{90}$ and AWF$_{100}._{500}$, respectively. In contrast, DeepSE-WF is able to process the AWF$_{100}._{4500}$ dataset within our memory requirements, and using a maximum of 34 GB during its operation. Figure~\ref{fig:estimationScaleComp} depicts the BER and MI estimations obtained by DeepSE-WF (and Tik-Tok's error) using the AWF$_{100}._{4500}$ dataset when compared to WFES and WeFDE in their previous setting. Figure~\ref{fig:estimationScaleComp} a) reveals that data constraints can severely affect the tightness of WF security estimations. For instance, for Front\_T1, the error of Tik-Tok is 0.16, and the BER estimate provided by WFES is 0.67. In turn, DeepSE-WF estimates a BER of $\approx$0.10. Surprisingly, neither the estimation for CS-BuFLO nor for Tamaraw have improved when using more data. These results allow us to make three observations. First, DeepSE-WF can generally provide tighter BER estimations when provided with additional data (e.g., it estimates a BER of $\approx$0.10 using AWF$_{100}._{4500}$ vs a BER of $\approx$0.48 using AWF$_{100}._{90}$ for the Front\_T1 defense). Second, additional data allows DeepSE-WF to produce estimations which lower bound the error obtained by existing state-of-the-art WF attacks like Tik-Tok, suggesting that even more accurate classifiers can be devised. Third, the BER estimation for constant-rate defenses seems to converge very fast (see also Section~\ref{sec:knn_convergence_eval}).

\begin{figure}[t]
\vspace{0.1cm}
    \centering
    \includegraphics[trim={0 0 0 2.4cm},clip,width=0.9\columnwidth]{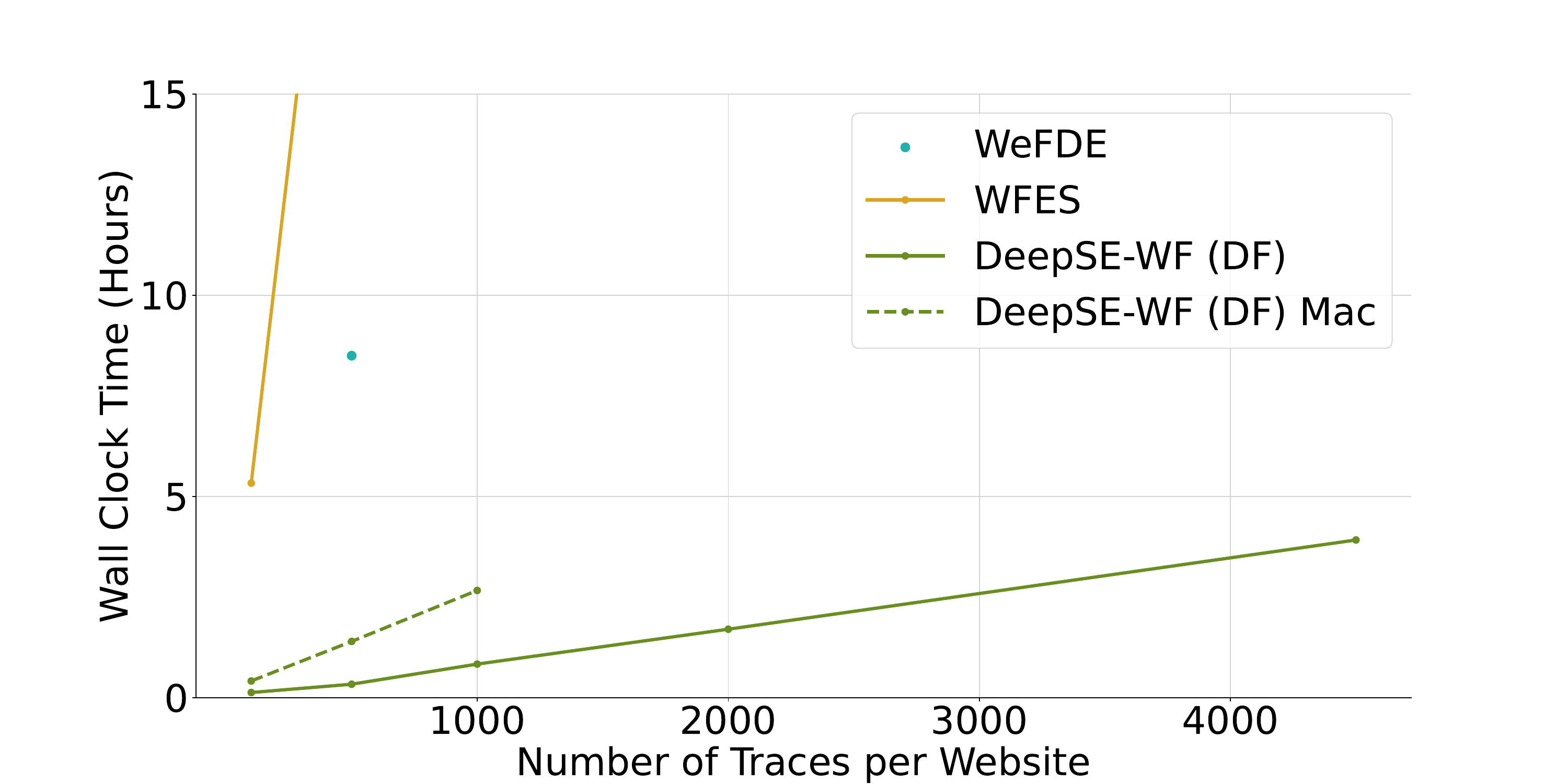}
    \vspace{-0.3cm}
    \caption{DeepSE-WF processing time for a single CV fold with increasing number of traces per website. WFES times are extrapolated based on a processing time of 3.7s/trace.}
    \label{fig:processing_speed}
    \vspace{-0.5cm}
\end{figure}

\mypara{DeepSE-WF is more efficient.} Our experiments have also revealed that, for the same problem scale, DeepSE-WF is able to deliver WF security estimation results much faster than existing approaches. Figure~\ref{fig:processing_speed} depicts the wall-clock time spent by DeepSE-WF, WFES, and WeFDE when producing estimations for different amounts of per-website samples. For instance, consider a security estimation over AWF$_{100}._{90}$, for a single defense, using our MacBook Pro testbed. In such a case, DeepSE-WF delivers its results within 25 min per cross-validation fold while WFES delivers its estimate after 9h 30 min for a single classifier estimation, over an order of magnitude slower. A similar case happens for the estimation of MI over AWF$_{100}._{500}$, where DeepSE-WF delivers its MI estimation within 1h 24 min and WeFDE delivers its estimate only after 8h 30 min. When we use the GPU server, DeepSE-WF can estimate BER and MI over AWF$_{100}._{4500}$ within 4h per cross-validation fold. Here the training time for both neural networks takes the most time with 3h 24 min, followed by the feature extraction which takes another 22 min. The final BER and MI estimation are significantly faster, only taking 4 and 7 min  respectively. The above results suggest that DeepSE-WF can scale to significantly larger datasets and obtain tighter security estimations when compared to existing security estimators. Extended results can be found in Appendix \ref{app:extended_results}.

\subsection{Convergence of the Estimations}
\label{sec:knn_convergence_eval}

Here, we evaluate the convergence rate of DeepSE-WF's estimates. Akin to Cherubin~\cite{wfes}, we study the asymptotic behaviour of BER in our finite AWF$_{100}._{4500}$ data, for an increasing amount of samples.

\begin{figure}[t]
    \centering
        \includegraphics[trim={0 0 0 2.4cm},clip,width=0.78\columnwidth]{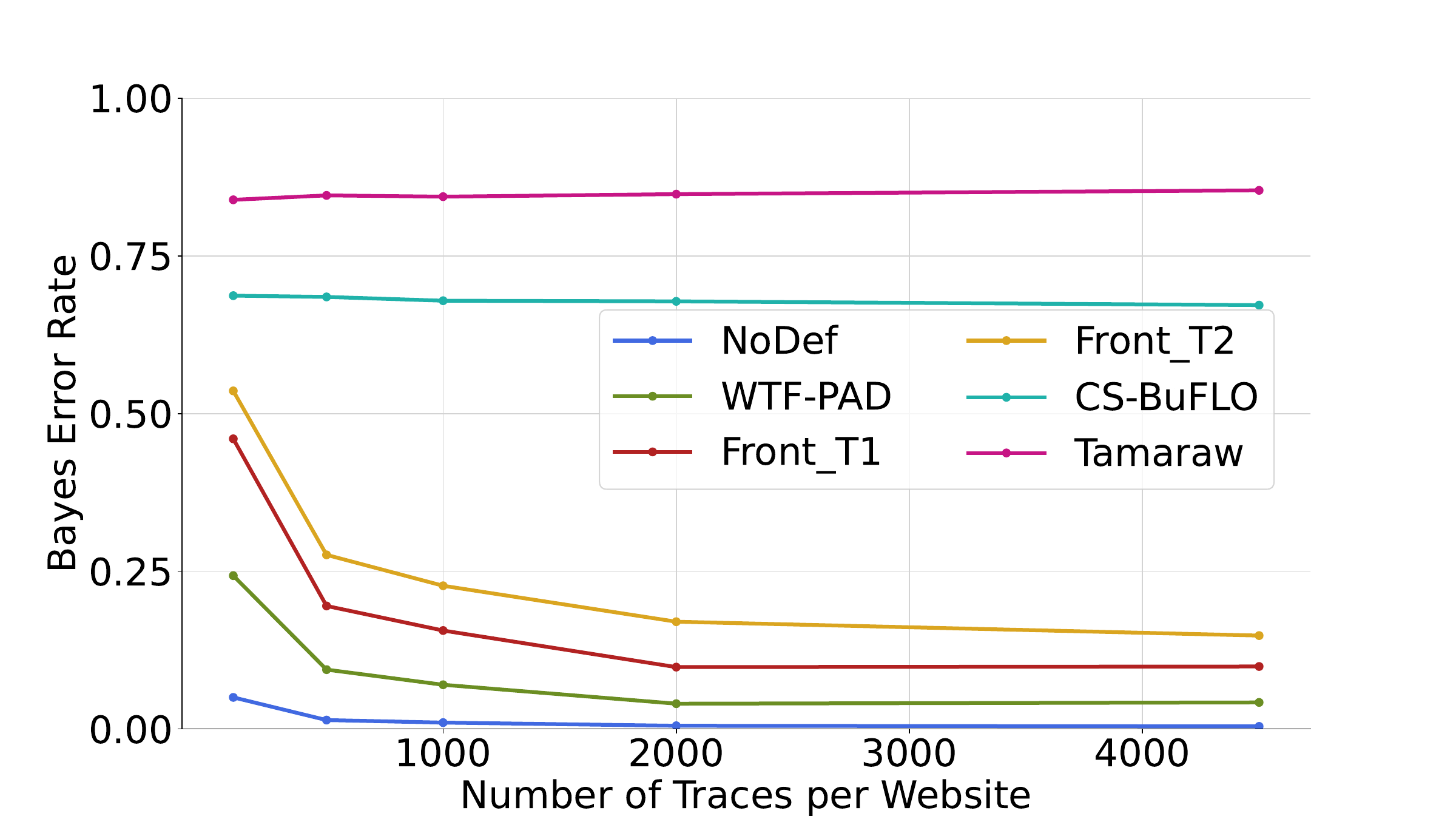}
    \vspace{-0.4cm}
    \caption{BER estimates for increasing traces per website.} 
         \vspace{-0.6cm}
    \label{fig:ber_scaling_samples}
\end{figure}

Figure~\ref{fig:ber_scaling_samples} depicts the evolution of DeepSE-WF's BER estimates, for multiple defenses, as the amount of samples available to the estimator increases (see Appendix~\ref{app:extended_results}).
The plot shows that DeepSE-WF reports an estimated BER below 20\% for all the defenses that try to obscure the original website trace by introducing noise (e.g., the Front variants), provided sufficient data ($\approx$ 2000 samples). This could mean that our estimation method is able to learn the randomly distributed noise of these defenses. In contrast, DeepSE-WF's BER estimation does not decrease as more data becomes available for constant rate defenses such as CS-BuFLO and Tamaraw. This suggests that DeepSE-WF converges to an estimation for the directional and timing representation with a low amount of samples. 

We also experiment with the importance of the testing set ($\mathcal{E}$) size for kNN, and observed that increasing the number of samples in $\mathcal{E}$ does not have a significant impact on the BER convergence. These results and the convergence of the MI can be found in Appendix~\ref{sec:extended_convergence}.

\subsection{Varying DeepSE-WF's DNN Architecture}
\label{sec:dnn_arch_params}

Our earlier experiments used the DF architecture to generate BER estimates. In this section, we explore the implications of using other DNN architectures proposed in previous WF attacks -- AWF-CNN, TF, and Var-CNN -- to generate BER estimates. We also extended these DNNs to use a direction + timing trace representation.

We note that, towards generalizing the WF attack across different scenarios and data distributions, TF is originally (pre-)trained on a set of websites and then fine-tuned and evaluated on a different set, making use of N-shot learning~\cite{triplet}. However, for allowing a direct comparison of TF's feature extractor with those of the other DNNs evaluated in our work, we train the TF architecture following the same methodology detailed in Section~\ref{sec:evaluation_setup}. Similarly to traditional transfer learning methods, developing a fair comparison using TF's original N-shot learning technique would entail additional work on understanding website samples' complexity in both upstream and downstream tasks. We relegate this task for future work.

\mypara{Training DeepSE-WF's DNNs.} We re-implemented the network architecture of deep learning-based WF attacks while using a learning rate of 0.002 with no weight decay and a batch size of 128. For each defense and trace representation, we train the network for 50 epochs with early stopping, and finally evaluate the classification error using the test set $\mathcal{T}$ before removing the last three layers.

The feature extractors of all DNN models are trained using binary cross-entropy loss, except for TF which originally uses either L2 or cosine loss. Throughout our evaluation, we use the former for training and estimating TF's BER since, for all other models, we estimate the BER by computing the k-NN error over latent features using the (Euclidean) L2 distance. In Appendix~\ref{app:tf}, we present further results on the use of TF when considering cosine loss, which provides tighter estimates vs. TF with L2 loss, but whose estimates are ultimately not as tight as the ones produced by DF or Var-CNN.

\begin{figure}[t]
    \centering
    \includegraphics[trim={0 0 0 2.4cm},clip,width=0.8\columnwidth]{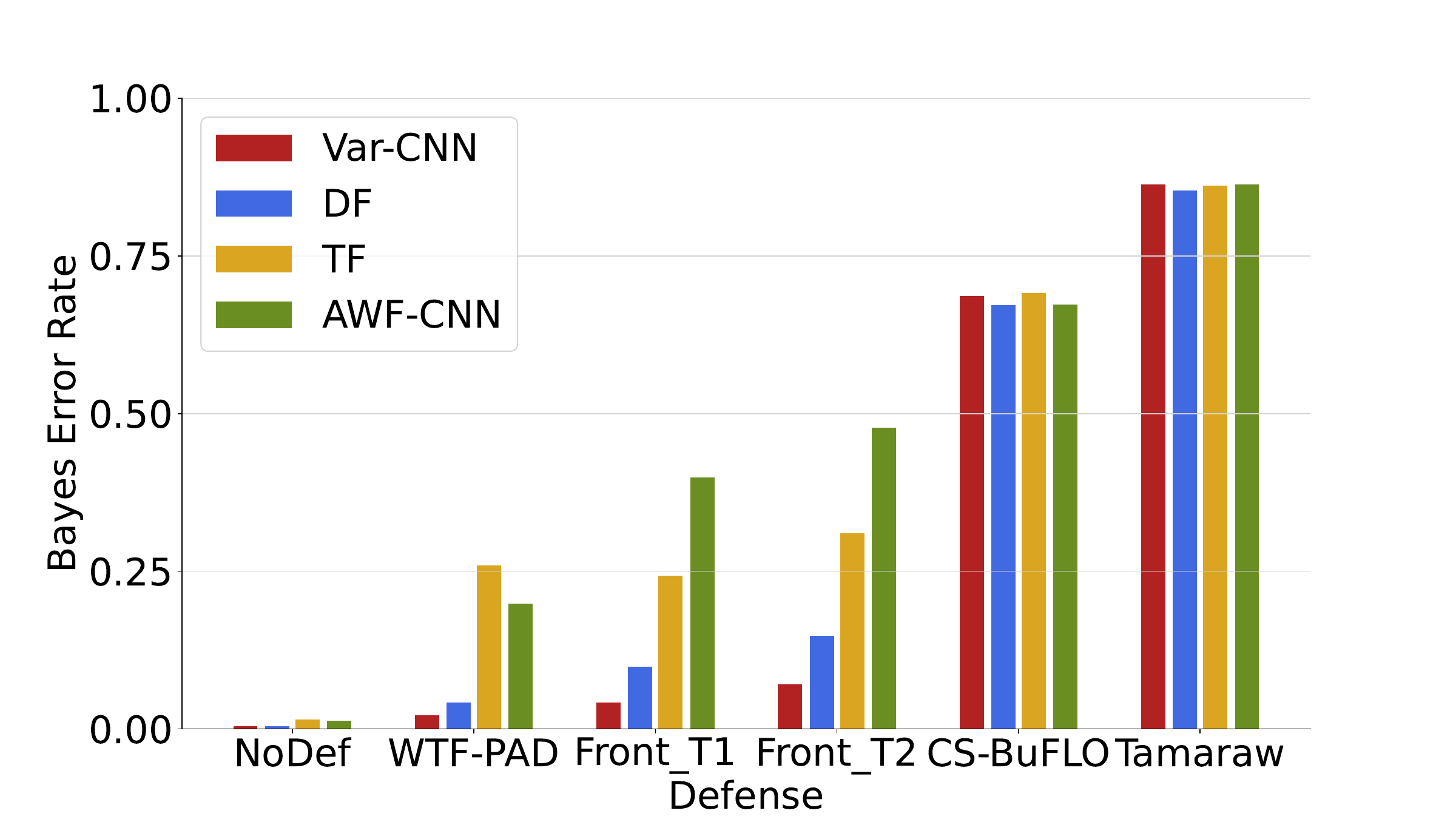}
    \vspace{-0.4cm}
    \caption{BER estimation using different DNN architectures, considering the AWF$_{100}._{4500}$ dataset.}
     \vspace{-0.4cm}
    \label{fig:ber_architecture_4500}
\end{figure}

\begin{figure}[t]
    \centering
        \includegraphics[trim={0 0 0 2.4cm},clip,width=0.8\columnwidth]{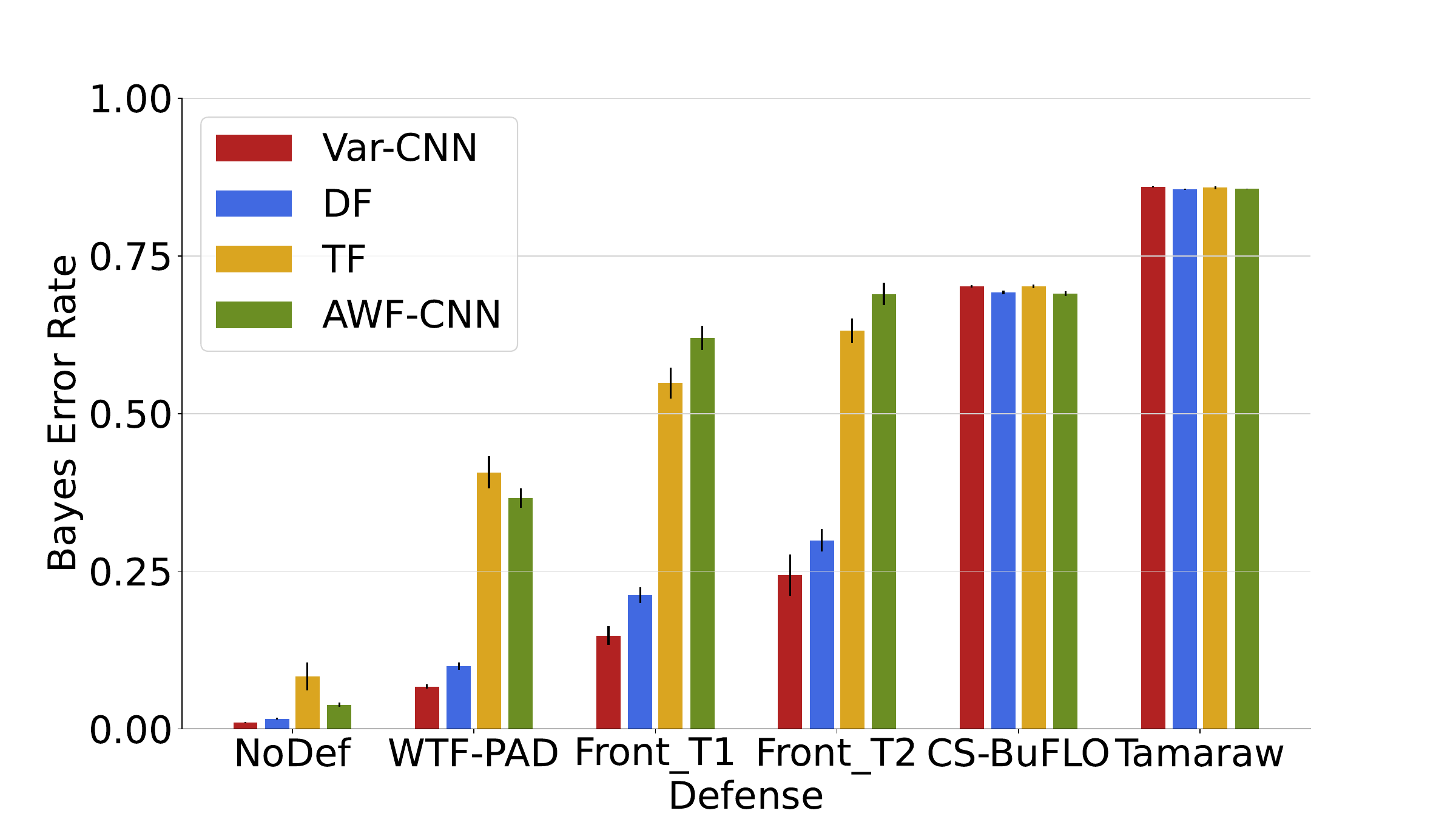}
    \vspace{-0.4cm}
    \caption{Mean and std.dev. BER estimation using different DNN hyperparameters, considering the AWF$_{100}._{500}$ dataset.}
     \vspace{-0.5cm}
    \label{fig:ber_hyperparam_4500}
\end{figure}

\mypara{Comparing the BER of different architectures.} Figure~\ref{fig:ber_architecture_4500} shows a bar plot comparison between the BER obtained by DeepSE-WF when estimating the security of WF defenses using different DNN architectures on the AWF$_{100}._{4500}$ dataset. 
We can see that, for constant rate defenses like CS-BuFLO and Tamaraw, the differences between the BER estimate obtained by different architectures is rather similar, varying only by a maximum of 1.9\%. 
For other defenses, we see that the choice of DNN architecture is a defining factor in the computation of tight BER estimates. For instance, for the Front\_T1 defense, we can observe that the AWF-CNN architecture obtains a BER of 39.9\%, more than double the BER obtained when using DF (9.9\%). Interestingly, the use of Var-CNN (a less recent architecture than TF or Tik-Tok) leads to tighter BER estimates for the defenses that do not produce constant-rate flows. 
The results of the above experiments show that the latent features produced by different DNN architectures have an impact on the estimation of the BER. Therefore, practitioners should strive to train DeepSE-WF with a selection of the most recent DNN architectures used in WF attacks to obtain the tightest security estimates for a particular defense. Ancillary experiments on the convergence behavior and processing time of different DNNs suggest that practitioners can also choose to trade-off shorter security estimate computation times for accurate (yet looser) security estimates (see Appendix~\ref{app:dnn_scale}). As an example, the BER estimates based on Var-CNN take about 3 times longer to complete when compared to the ones based on the DF architecture, for a decrease of 5.7\% in  the BER bound for Front\_T1.

\mypara{Optimization of DNNs hyperparameters.} To assess whether DeepSE-WF's results are robust with regard to the parameterization of the DNN architecture used for estimation, we conducted a sensitivity study to gauge the tightness of DeepSE-WF BER bounds when changing a selection of the considered DNNs' hyperparameters. In our experiments, we chose different configurations for \textit{dropout} and \textit{embedding\_size}, representing the most important aspect for regularization of neural networks and thus restricting their overfitting capability. We prevent the need for tuning hyperparameters of the optimizer by fixing a small learning rate with many epochs, and enable early stopping for efficiency.
The results of our experiments, depicted in Figure~\ref{fig:ber_hyperparam_4500}, show the mean and standard deviation of the BER over all the different combinations of $dropout \in (0.1, 0.2, 0.3)$, and $embedding\_size \in (64, 128, 512)$ in each of the DNN architectures. We can observe that the BER estimates produced by each DNN remain rather stable, with a maximum standard deviation of 3.3\% for the Var-CNN DNN when considering the Front\_T2 defense.

\mypara{DeepSE-WF's BER estimates vs. current WF attacks' error.} Table~\ref{table:attackAndBoundEval} compares DeepSE-WF's BER estimation against the classification error obtained by well-known WF attacks on a selection of relevant WF defenses.
The table reveals that DeepSE-WF consistently obtains BER estimations below the classification error of existing WF attacks. For instance, DeepSE-WF (Var-CNN) obtains an average BER estimate of 7.1\% for Front\_T2 while the Var-CNN attack implementation, i.e., the best performing attack against this defense, is only able to reach an average classification error of 11.1\%. This observation is more pronounced for constant-rate padding defenses, where k-FP actually performed better than existing deep learning-based attacks. For instance, for CS-BuFLO, DeepSE-WF (DF) obtains an average BER estimate of 67.2\% while the k-FP attack achieves a classification error of 80.9\%. 
These results suggest that WF attacks have a wide margin for further improvements.

\subsection{Experiments on the DS19 Dataset}
\label{sec:ds19_dataset}

So far, our experimental evaluation leveraged the widely available AWF~\cite{AWF} dataset, which was collected in 2017. Since then, 
there have been a number of improvements to Tor (e.g., padding schemes to protect against traffic correlation~\cite{tor4alpha}). To understand whether more recent versions of Tor provide perceivable improvements against WF, and whether DeepSE-WF security estimations hold when using a more recent dataset, we used the DS19 dataset~\cite{front}, collected in 2019, to perform an additional validation of our results.

We leveraged the DS19 dataset to perform a set of experiments that followed the same methodology used throughout this section.
Table~\ref{table:AWFToDS19Results} and Table~\ref{table:ds19Results} depict the BER estimations of different DNN architectures when considering the AWF$_{100}._{100}$ and DS19$_{100}._{100}$ datasets, respectively. To perform a direct comparison between these datasets, we used a sample of AWF with 100 traces per website to match the number of traces made available in the DS19 dataset. For the considered datasets, DeepSE-WF's BER estimates reveal a similar trend to our previous experiments: the BER of constant-rate padding defenses is typically larger than the error achieved for other defenses, while the estimates using the DF and Var-CNN architectures generally outperform the AWF-CNN and TF ones.

\begin{table}[t]
\centering
\caption{Classification error (in \%) for WF attacks on WF defenses compared to our BER estimation on AWF$_{100}._{4500}$.}
\vspace{-0.4cm}
\resizebox{\columnwidth}{!}{%
    \begin{tabular}{lcccccc}
        \toprule
        \textbf{Attacks \& Estimators}       & \textbf{NoDef}                & \textbf{WTF-PAD}                 & \textbf{Front\_T1}               & \textbf{Front\_T2}               & \textbf{CS-BuFLO}                & \textbf{Tamaraw} \\ \midrule
        {k-FP}                               & 04.1 $\pm$ 0.0                & 33.0 $\pm$ 0.0                   & 41.2 $\pm$ 0.2                   & 46.3 $\pm$ 0.1                   & 80.9 $\pm$ 0.1                   & 93.2 $\pm$ 0.1   \\
        {AWF-CNN}                            & 03.5 $\pm$ 0.1                & 37.5 $\pm$ 0.9                   & 51.0 $\pm$ 0.5                   & 60.7 $\pm$ 0.4                   & 84.6 $\pm$ 0.5                   & 94.9 $\pm$ 0.1   \\
        {DF}                                 & 00.7 $\pm$ 0.0                & 07.4 $\pm$ 0.1                   & 15.8 $\pm$ 0.1                   & 22.9 $\pm$ 0.1                   & 83.0 $\pm$ 0.1                   & 94.8 $\pm$ 0.1   \\
        {TF (L2 loss)}                                 & 02.9 $\pm$ 0.4 & 45.4 $\pm$ 2.0   & 42.6 $\pm$ 2.1   & 52.2 $\pm$ 4.8   & 90.0 $\pm$ 0.1   & 97.3 $\pm$ 0.3 \\
        {Var-CNN}                            & 00.7 $\pm$ 0.1                & 03.3 $\pm$ 0.1                   & 06.4 $\pm$ 0.2                   & 11.1 $\pm$ 1.3                   & 83.0 $\pm$ 0.0                   & 96.0 $\pm$ 2.0   \\
        {Tik-Tok}                            & 01.0 $\pm$ 0.1                & 06.5 $\pm$ 0.2                   & 15.9 $\pm$ 0.6                   & 22.3 $\pm$ 0.2                   & 82.8 $\pm$ 0.1                   & 94.8 $\pm$ 0.1   \\
    \midrule
        {DeepSE-WF (AWF-CNN)} & 01.3 $\pm$ 0.1 & 19.9 $\pm$ 0.2 & 39.9 $\pm$ 0.2 & 47.8 $\pm$ 0.5  & 67.3 $\pm$ 0.1 & 86.3 $\pm$ 1.1 \\
        {DeepSE-WF (DF)}      & \textbf{00.4 $\pm$ 0.0}                    & 04.2 $\pm$ 0.2                  & 09.9 $\pm$ 0.2                   & 14.8 $\pm$ 0.2                    & \textbf{67.2 $\pm$ 0.1}                   & \textbf{85.4 $\pm$ 1.1}  \\
        {DeepSE-WF (TF - L2 loss)}      & 01.5 $\pm$ 0.2 & 25.9 $\pm$ 1.3 & 24.3 $\pm$ 1.4 & 31.0 $\pm$ 3.7 & 69.1 $\pm$ 0.2 & 86.1 $\pm$ 1.2 \\
        {DeepSE-WF (Var-CNN)} & \textbf{00.4 $\pm$ 0.0} & \textbf{02.2 $\pm$ 0.1} & \textbf{04.2 $\pm$ 0.1} & \textbf{07.1 $\pm$ 0.2}  & 68.6 $\pm$ 0.5 & 86.3 $\pm$ 1.1 \\ \bottomrule
        \end{tabular}
}
    \label{table:attackAndBoundEval}
    \vspace{-0.45cm}
\end{table}

Tables~\ref{table:AWFToDS19Results} and~\ref{table:ds19Results} also reveal a discrepancy between the BER estimates obtained by the same DNN when considering the different datasets. Several factors may contribute to these differences~\cite{WFcritics,onlineWF}. First, the version of Tor used to collect DS19 incorporates multiple updates which can translate in changes to websites’ traffic patterns. Second, since the top 100 websites included in each dataset were drawn from the (dynamic) Alexa Top Sites list, these websites may not be the same in both datasets. Third, even if a fraction of websites are the same in both datasets, it is plausible that their content (and their corresponding traffic signatures) have changed between the collection of AWF (2017) and DS19 (2019). Hardware/network configurations of traffic collectors may lead to further discrepancies.

In addition, our experiments also help showcasing differences in the estimates produced by different DNN architectures depending on the amounts of available data. For instance, DF shows better results than VarCNN in Table 4 (when using AWF$_{100}._{100}$), as opposed to Table 3 (when using AWF$_{100}._{4500}$), suggesting that VarCNN can obtain tighter estimates at the cost of a larger training dataset.

\section{Theoretical Analysis}

We now discuss provable bounds for DeepSE-WF estimates (Section~\ref{sec:merged_eval}) and the relationship between BER and MI (Section~\ref{subsec:ber_vs_mi}).

\subsection{Evaluation of Provable Bounds}
\label{sec:merged_eval}

Most WF defenses do not have provable bounds on the maximum accuracy possible for any attack on them. We now introduce a simulated defense based on \textit{merged traces}, where a user combines the traces of $M$ websites loaded at the same time. While such a defense implies a large overhead and would thus unlikely be considered for hardening Tor against WF attacks in practical settings, it provides us with compelling theoretical properties that enable us to evaluate provable bounds of DeepSE-WF's estimations -- such a defense has the provable property that the BER is exactly $1 - \frac{1}{M}$.

\begin{table}[t!]
\centering
\caption{BER estimation with different DNNs on AWF$_{100}._{100}$.}
\vspace{-0.4cm}
\resizebox{0.95\columnwidth}{!}{%
    \begin{tabular}{lcccccc}
        \toprule
        \textbf{Architecture} & \textbf{NoDef} & \textbf{WTF-PAD} & \textbf{Front\_T1} & \textbf{Front\_T2} & \textbf{CS-BuFLO} & \textbf{Tamaraw} \\ \midrule
        AWF-CNN & 09.2 $\pm$ 0.2  & 50.4 $\pm$ 3.0   & 74.5 $\pm$ 2.7     & 77.9 $\pm$ 2.0     & 70.4 $\pm$ 1.4    & 84.8 $\pm$ 1.1   \\
        DF      & 05.0 $\pm$ 0.3  & 24.3 $\pm$ 0.7   & 46.0 $\pm$ 0.4     & 53.6 $\pm$ 1.3     & 68.7 $\pm$ 0.7    & 83.9 $\pm$ 0.6   \\
        {TF (L2 loss)}    & 25.3 $\pm$ 1.5 & 56.4 $\pm$ 1.8 & 73.1 $\pm$ 1.2 & 77.6 $\pm$ 0.8 & 71.3 $\pm$ 0.6 & 84.9 $\pm$ 0.9 \\
        Var-CNN & 03.0 $\pm$ 0.3  & 29.1 $\pm$ 9.2   & 50.9 $\pm$ 2.3     & 61.8 $\pm$ 2.3     & 73.1 $\pm$ 0.9    & 84.3 $\pm$ 0.8   \\ \bottomrule
    \end{tabular}
}
    \label{table:AWFToDS19Results}
    \vspace{-0.3cm}
\end{table}

\begin{table}[t!]
\centering
\caption{BER estimation with different DNNs on DS19$_{100}._{100}$.}
\vspace{-0.4cm}
\resizebox{0.95\columnwidth}{!}{%
    \begin{tabular}{lcccccc}
        \toprule
        \textbf{Architecture} & \textbf{NoDef}   & \textbf{WTF-PAD} & \textbf{Front\_T1} & \textbf{Front\_T2} & \textbf{CS-BuFLO} & \textbf{Tamaraw} \\ \midrule
        AWF-CNN & 05.9 $\pm$ 0.3  & 34.3 $\pm$ 1.9  & 68.8 $\pm$ 1.2  & 77.7 $\pm$ 2.3  & 68.0 $\pm$ 1.7  & 74.3 $\pm$ 1.4 \\
        DF      & 02.8 $\pm$ 0.2  & 07.3 $\pm$ 0.5  & 18.8 $\pm$ 0.5  & 30.9 $\pm$ 1.2  & 67.2 $\pm$ 0.6  & 72.5 $\pm$ 1.5 \\
        {TF (L2 loss)}    & 12.6 $\pm$ 1.0 & 30.1 $\pm$ 0.5 & 61.5 $\pm$ 1.4 & 71.4 $\pm$ 1.6 & 67.9 $\pm$ 0.7 & 73.7 $\pm$ 1.3 \\
        Var-CNN & 03.1 $\pm$ 1.0  & 06.7 $\pm$ 0.4  & 20.7 $\pm$ 4.8  & 36.4 $\pm$ 5.5  & 72.3 $\pm$ 1.0  & 72.1 $\pm$ 1.2 \\ \bottomrule
    \end{tabular}
}
    \label{table:ds19Results}
    \vspace{-0.4cm}
\end{table}

\mypara{Merged Traces.} To simulate a scenario where a user can simultaneously load $M$ pages, we concatenate $M-1$ randomly chosen traces from our dataset with a pre-selected trace and then sort all packets on the traces by their respective timestamps. In such a setting, an adversary can at best randomly guess which website is the pre-selected one, even in the unlikely event that she can distinguish all $M$ traces. With that, the adversary has at most an expected success of $\frac{1}{M}$, where we assume that the theoretical error for $M=1$ is 0 (i.e., we can perfectly classify the undefended traces).

\begin{figure}[t]
    \centering
        \includegraphics[trim={0 0 0 2.4cm},clip,width=0.36\textwidth]{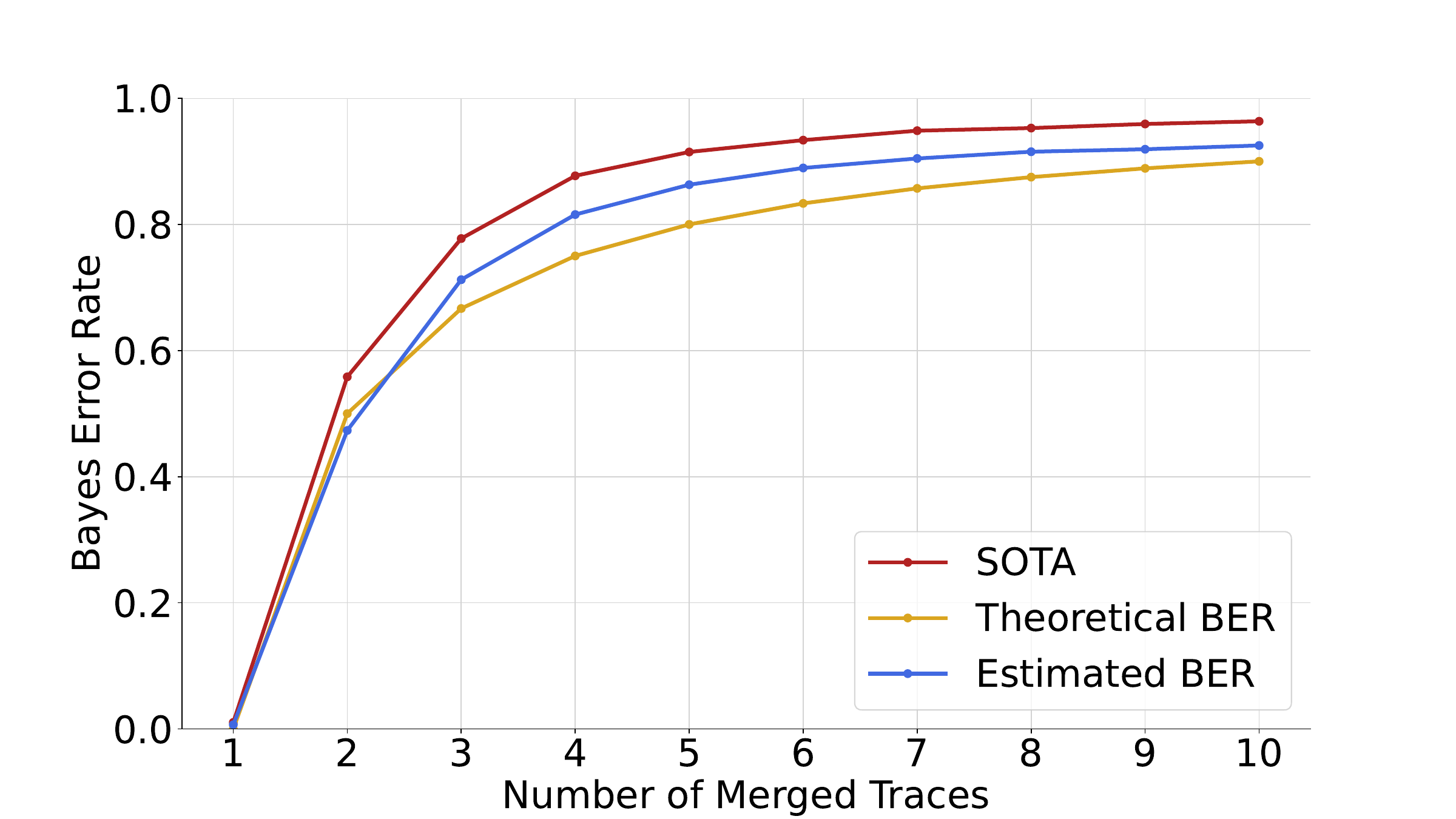}
        \vspace{-0.3cm}
        \caption{Comparison of our estimated BER with Tik-Tok and the theoretical BER using the merged traces.}
        \vspace{-0.4cm}
        \label{fig:merged_ber}
\end{figure}

\mypara{Evaluation.} We simulate the above scenario for $M \in [1, 10]$ and set a baseline with the Tik-Tok attack~\cite{Tik-Tok} and the AWF$_{100}._{1000}$ dataset. The classification error of that attack will serve as an upper bound of the BER. The BER is estimated using DeepSE-WF (with Tik-Tok's DNN) on the same dataset.
Figure~\ref{fig:merged_ber} compares our estimation and Tik-Tok to the theoretical BER. The plot shows that for all $M$, we lower-bound the Tik-Tok attack but due to the finite amount of data and feature transformations (c.f.  Section~\ref{subsec:reasoning}), we are not able to provide a lower bound of the true BER. This could also be due to the fact that correctly identifying the $M$ websites from a single trace is a  hard problem. Another reason could be the sheer dimension of the traces -- merging up to 10 traces into a single one can increase the overall length above 50000. Truncating the trace to a length of 5000 might ignore useful information, increasing the BER.

\subsection{Comparing the BER and MI}
\label{subsec:ber_vs_mi}

\begin{figure}[t]
    \centering
    \begin{subfigure}[t]{0.23\textwidth}
        \includegraphics[trim={3cm 0 4.5cm 2.4cm},clip,width=\textwidth]{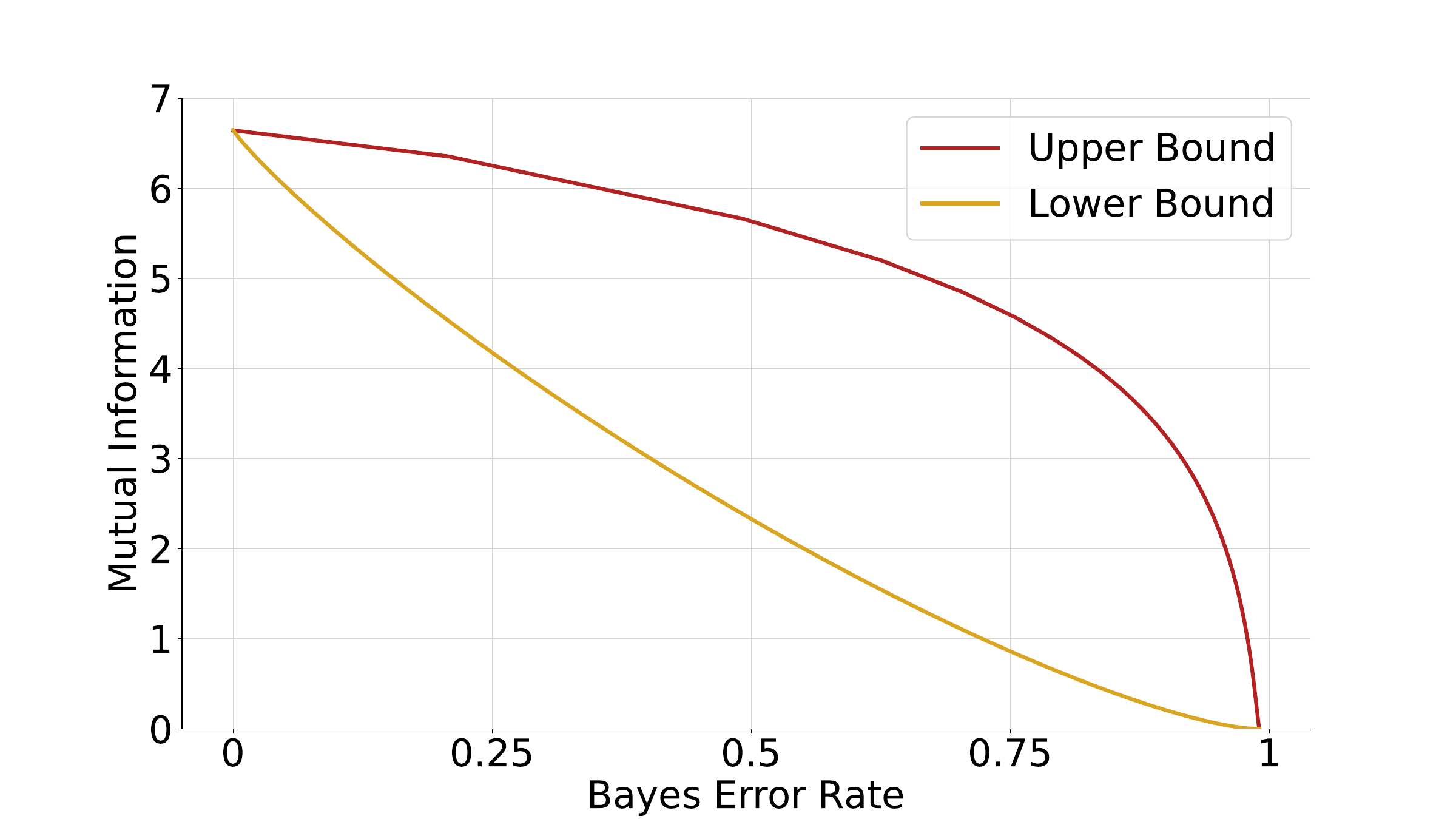}
        \vspace{-0.6cm}
        \caption{}
        \label{fig:ber_mi_bounds}
    \end{subfigure}
    \begin{subfigure}[t]{0.23\textwidth}
        \includegraphics[trim={2cm 0 4.5cm 2.4cm},clip,width=\textwidth]{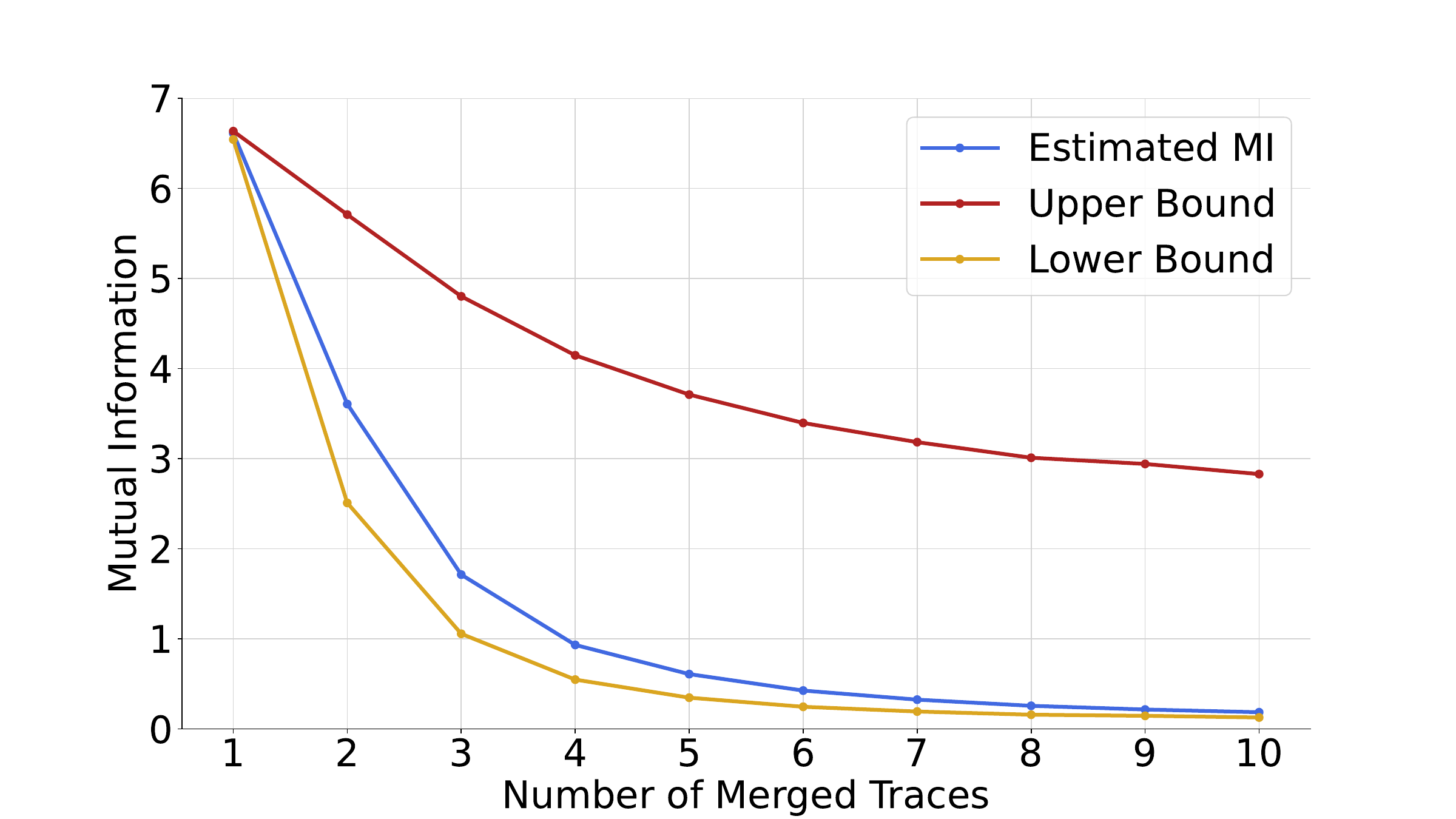}
        \vspace{-0.6cm}
        \caption{}
        \label{fig:merged_mi}
    \end{subfigure}
    \vspace{-0.4cm}
     \caption{Placement of DeepSE-WF MI estimations (b) within the Fano and Kovalevskij bounds for MI (a).}
     \vspace{-0.6cm}
    \label{fig:ber_mi_comparison}
\end{figure}

In this section, we show why using estimators of the mutual information and the Bayes error rate result in similar insights.
Although Li et. al.~\cite{wefde} show that for \textit{some} classifiers with very small accuracy (different from the Bayes classifier), there exists a distribution yielding high information leakage (c.f., Figure~1 and Theorem~1 in~\cite{wefde}), we explain why this is not the case when relating the BER to MI.
We achieve this by relying on well-established information theoretical bounds between the BER and MI~\cite{fano1961transmission, hellman1970probability, kovalevsky1968problem}. To understand the difference between our analysis and Theorem~1 by Li et. al.~\cite{wefde}, note that the accuracy of the Bayes classifier (1 - the BER) can not be smaller than $1/n$ for classifying $n$ different websites. For this, the posterior distribution must be equal to a uniform distribution, as the Bayes classifier would have a strictly larger accuracy otherwise. This aspect is violated in the constructed example for the proof of Theorem~1 by Li et. al.~\cite{wefde}, suggesting that the upper bounds derived there are not tight for the Bayes classifier accuracy.
Using Fano's bound~\cite{fano1961transmission}, we can lower bound the MI given the BER by:
\begin{equation}\label{Fano_mi}
\begin{small}
I(W;T) \geq H(|W|)- H(R^*) - R^*\log_2(|W|-1),
\end{small}
\end{equation}
where $R^*$ is the BER, $W$ is the set of websites and $T$ the traces. Likewise we can upper bound the MI with Kovalevskij's~\cite{kovalevsky1968problem} bounds:
\begin{align}\label{kovalevskij}
\begin{split}
I(W;T) \leq & \min_{2\leq k \leq |W|} \ H(|W|) - \log_2(k)  \\
&- k(k+1)\log_2\left(\frac{k+1}{k}\right)\left(R^* -\frac{k-1}{k} \right),
\end{split}
\end{align}
where again $W$ is the set of websites. Fano's~\cite{fano1961transmission} and Kovalevskij's~\cite{kovalevsky1968problem} bounds are shown in Figure \ref{fig:ber_mi_comparison} a). To understand whether DeepSE-WF produces reasonable MI estimations, we plug in the theoretical BER values for the merged traces into the Fano and Kovalevskij bounds to obtain upper and lower bounds for the true MI of our merged traces. This exercise, depicted in Figure \ref{fig:ber_mi_comparison} b), reveals that our estimation is in the range of possible MI values, while being located close to the lower bound.

\section{Conclusions}
\label{sec:conclusions}

In this paper, we showed that current security estimation frameworks overestimate the protection offered by existing WF defenses against deep learning WF attacks. To tackle this issue, we proposed DeepSE-WF, a novel method for estimating the security of WF defenses based on Nearest Neighbors Bayes error and mutual information estimators deployed over latent feature spaces.

While disregarding the need to manually craft features, DeepSE-WF is not WF attack agnostic. Instead, DeepSE-WF picks one amongst several existing DNN architectures (used in current attacks) and corresponding hyperparameters that yield the best BER/MI estimate. This methodology can be perceived as an estimate of the potential of a given deep learning approach. Indeed, the ML literature~\cite{renggli2022model} showed that a similar technique can serve as a proxy for model selection in transfer learning and the visual domain, but this would require further investigation in the context of WF.

In addition, since DeepSE-WF simply picks the DNN representation that yields the lowest BER (or higher MI) estimate, it does not provide or require further comparisons between learned representations. Orthogonal to our work, the ML community has been conducting a flurry of work aimed at comparing (e.g., via similarity metrics) and understanding neural network representations, ranging from early work like centered kernel alignment (CKA)~\cite{kornblith2019similarity}, to modern approaches like model stitching~\cite{lenc2015understanding, bansal2021revisiting, csiszarik2021similarity}. 

Despite its reliance in known attacks, DeepSE-WF offers compelling opportunities to improve the design and evaluation process of WF defenses by a) allowing researchers to evaluate WF defenses against the potential exhibited by state-of-the-art DNN-based attacks, and b) prompting future research on the effectiveness of WF defenses to analyse sample complexity and convergence. From our theoretical understanding, we know to which value a BER or MI estimate should converge (exactly for synthetic traces).

\mypara{Limitations and future directions.} Our work has laid out the groundwork for a more comprehensive set of experiments on the examination of WF defenses' security estimation methods.

First, in contrast to WFES and WeFDE, our method analyses latent feature spaces and is unable to provide human-interpretable data about the most informative traffic features (i.e., which traffic features should be better hidden when developing new defenses). Nevertheless, we stress that our method does not ignore information from the defended traces' representations, and that current research trends on explainable deep learning~\cite{lime, shap} may allow DeepSE-WF to produce interpretable feature analyses in the future.

Second, DeepSE-WF does not provide security estimations for defenses in the open-world setting due to the lack of exact number of classes and known prior website probabilities in this setting. The latter is crucial for formulating information theoretical quantities such as MI or the BER. Nonetheless, as stressed out in Section~\ref{sec:wf_background}, assuming a uniform probability for websites in a closed-world setting yields a lower bound for the security of any defense mechanism. Given this fact, we are, by comparing lower bound estimates, unable to assess, for instance, whether two defenses that provide rather disparate security guarantees in the closed-world setting can offer similar security in the open world setting. An interesting direction for future work would be to adapt DeepSE-WF to enable security estimations in the open world setting.

\bibliographystyle{ACM-Reference-Format}
\bibliography{References}

\appendix

\appendix
\def\thesection{}
\def\thesubsection{\Alph{subsection}}
\def\thesubsubsection{\thesection.\arabic{subsubsection}}
\section{\Kern{-1em}Appendix}

\subsection{Alternate MI Estimators}
\label{app:mine}

When building DeepSE-WF, we have also experimented with MINE~\cite{mine}, another recently proposed method for estimating the MI of high dimensional variables. 
Similar to our work, MINE requires the training of neural networks to generate learned features used in estimations. However, MINE may unpredictably over- or under-estimate MI due to its sensitivity to hyperparameters and to the use of particular neural network structures~\cite{mine}. Thus, while promising, we relegate a comprehensive study of the ability of MINE to accurately estimate the security of WF defenses for future work.

\begin{figure}[b]
    \centering
        \includegraphics[trim={0 0 0 2.4cm},clip,width=0.45\textwidth]{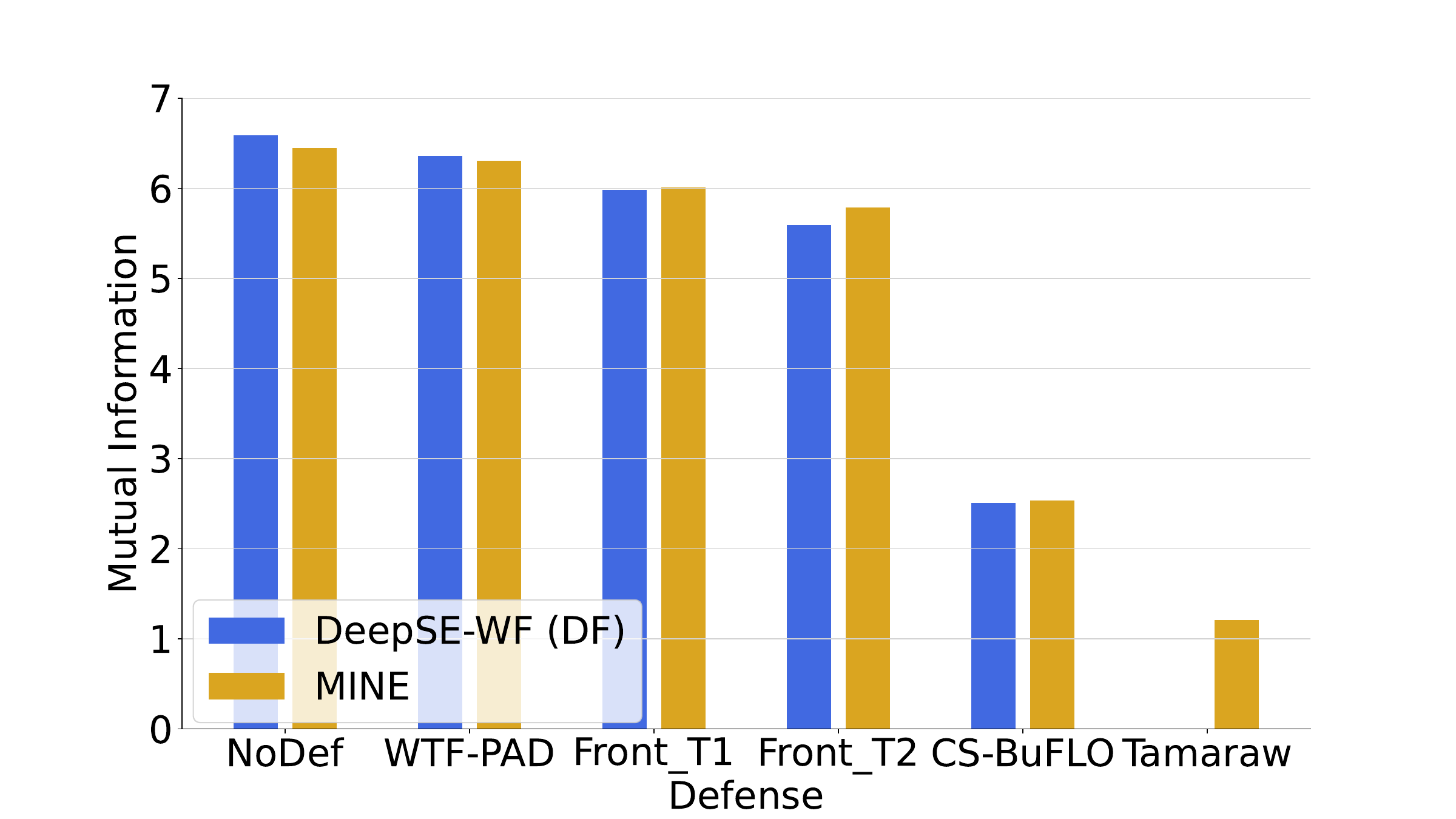}
        \vspace{-0.3cm}
        \caption{Comparison of DeepSE-WF and MINE MI bounds (AWF$_{100}._{4500}$).}
        \vspace{-0.4cm}
        \label{fig:mine}
\end{figure}

Nevertheless, we deliver a preliminary comparison of DeepSE-WF and MINE, by selecting a set of MINE hyperparameters that maximize MI for our (non-) defended data. As we see in Figure~\ref{fig:mine}, both approaches produce reasonably close estimations. Therefore, we pragmatically chose to use a kNN-based MI estimation in DeepSE-WF since we can use the same learned features both for estimating the BER and MI, and our approach does not depend on hyperparameter tuning (apart from that already performed in the original WF attacks' DNNs we used to learn latent features).

\subsection{MI and Feature Transformations}
\label{app:proof_mi_transformations}

We prove the following statement made in the main body of this work:

\begin{lem}
\label{lem:MI_Transformations}
For any deterministic transformation $f$ and two random variables $X$ and $Y$, where a realization of $x \sim X$ represent the features of a sample and $y \sim Y$ its label, it holds that $I(X;Y) \geq I(f(X);Y)$, where $I(\cdot;\cdot)$ represent the mutual information.
\end{lem}

\begin{proofof}{Lemma \ref{lem:MI_Transformations}}
Notice that $Y \rightarrow X \rightarrow f(X)$ for a Markov chain, i.e. $f(X)$ and $Y$ are independent given $X$. Applying the \textit{data processing inequality}~\cite{cover1991information}, it directly follows that $I(Y;f(X)) \leq I(Y;X)$. Symmetry of the mutual information concludes the proof.
\end{proofof}

\subsection{Defending the Dataset}
\label{sec:hyperparams}

To evaluate our estimation approaches and assess their performance, we used a number of defenses which are simulated on the timing representation. While we use default parameters for WF defenses, as suggested in their original papers, here we shed light on the particular configurations used.

We used the WTF-PAD~\cite{wtfpad} implementation included in the repository of WFES~\cite{wfes}, and two versions of Front \cite{front} with parameters $N_c = N_s = 1700$, $W_{min}=1$ and $W_{max} =14$ for Front\_T1 and $Nc =Ns =2500$ for Front\_T2. Since Front\_T2 has a larger sampling window, it should induce more dummy packets to the trace and therefore have a higher BER than Front\_T1. For CS-BuFLO~\cite{csbuflo} and Tamaraw~\cite{tamaraw}, we use the parameters $d = 1$ and $2^{-4} * 1000 \leq \rho \leq 2^3 * 1000$ and $\rho_{out} = 0.04$, $\rho_{in} = 0.012$ with $L = 50$, respectively.

\subsection{Extended Results}
\label{app:extended_results}

This section offers a more comprehensive numerical outlook on the estimations provided by DeepSE-WF and past WF security estimation frameworks.

\begin{figure}[t]
    \centering
        \includegraphics[trim={0 0 0 2.4cm},clip,width=\columnwidth]{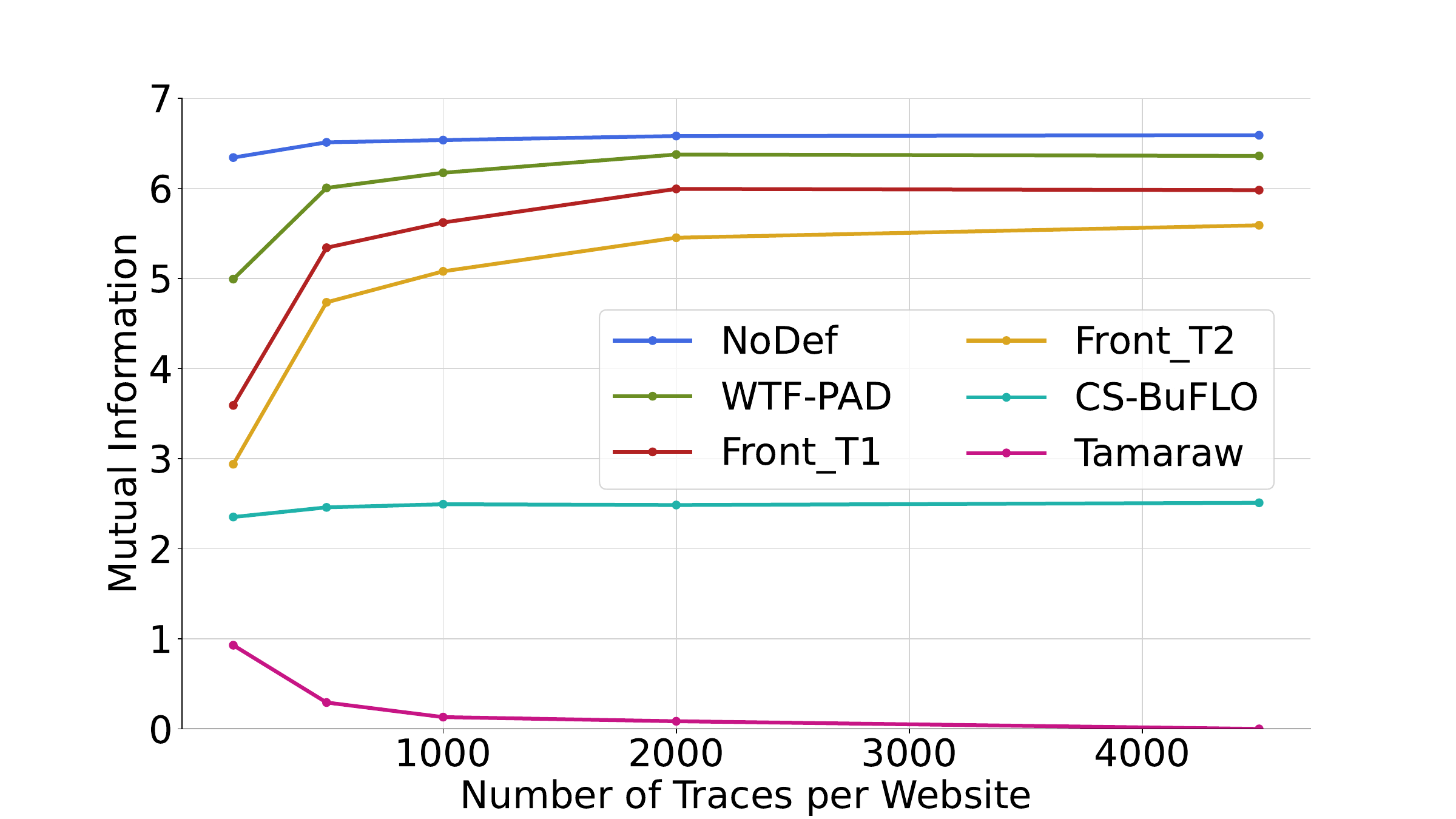}
        \caption{DeepSE-WF MI estimation for increasing number of traces per website (AWF$_{100}._{4500}$).}
        \label{fig:mi_convergence}
        \vspace{-0.3cm}
\end{figure}

\mypara{DeepSE-WF estimates.} Table~\ref{tab:full_ber} includes the final BER (in \%) estimated by DeepSE-WF for an increasing number of samples per website, including the standard deviation of our 5-fold cross-validation. Table~\ref{tab:full_mi} includes the final MI (in bits) estimated by DeepSE-WF in the same conditions. We show these results in addition to the graphical overview shown in Figure~\ref{fig:ber_scaling_samples} and Figure~\ref{fig:mi_convergence}.

\mypara{WFES Bayes Error Rate estimates.} Table~\ref{tab:full_wfes} shows the BER estimates produce by WFES (over the AWF$_{100}._{90}$ dataset) for different feature sets considered by well-known WF attacks based on manual feature engineering. We can observe that WFES obtains the majority of its smaller bounds estimates when using the k-FP attack feature set, while generally overestimating the results obtained by DeepSE-WF, even for the same amount of samples (refer to column \texttt{90} in Table~\ref{tab:full_ber}).

\mypara{WeFDE and MINE mutual information estimates.} Table~\ref{table:MI_results} shows the WeFDE and MINE mutual information estimates over the AWF$_{100}._{500}$ and AWF$_{100}._{4500}$ datasets, respectively. The numbers in this table suggest that WeFDE overestimates MI when compared to MINE, and that MINE obtains similar results to DeepSE-WF's kNN-based MI estimates (refer to the rightmost column of Table~\ref{tab:full_mi}).

\begin{figure}[t]
    \centering
        \includegraphics[trim={0 0 0 2.4cm},clip,width=\columnwidth]{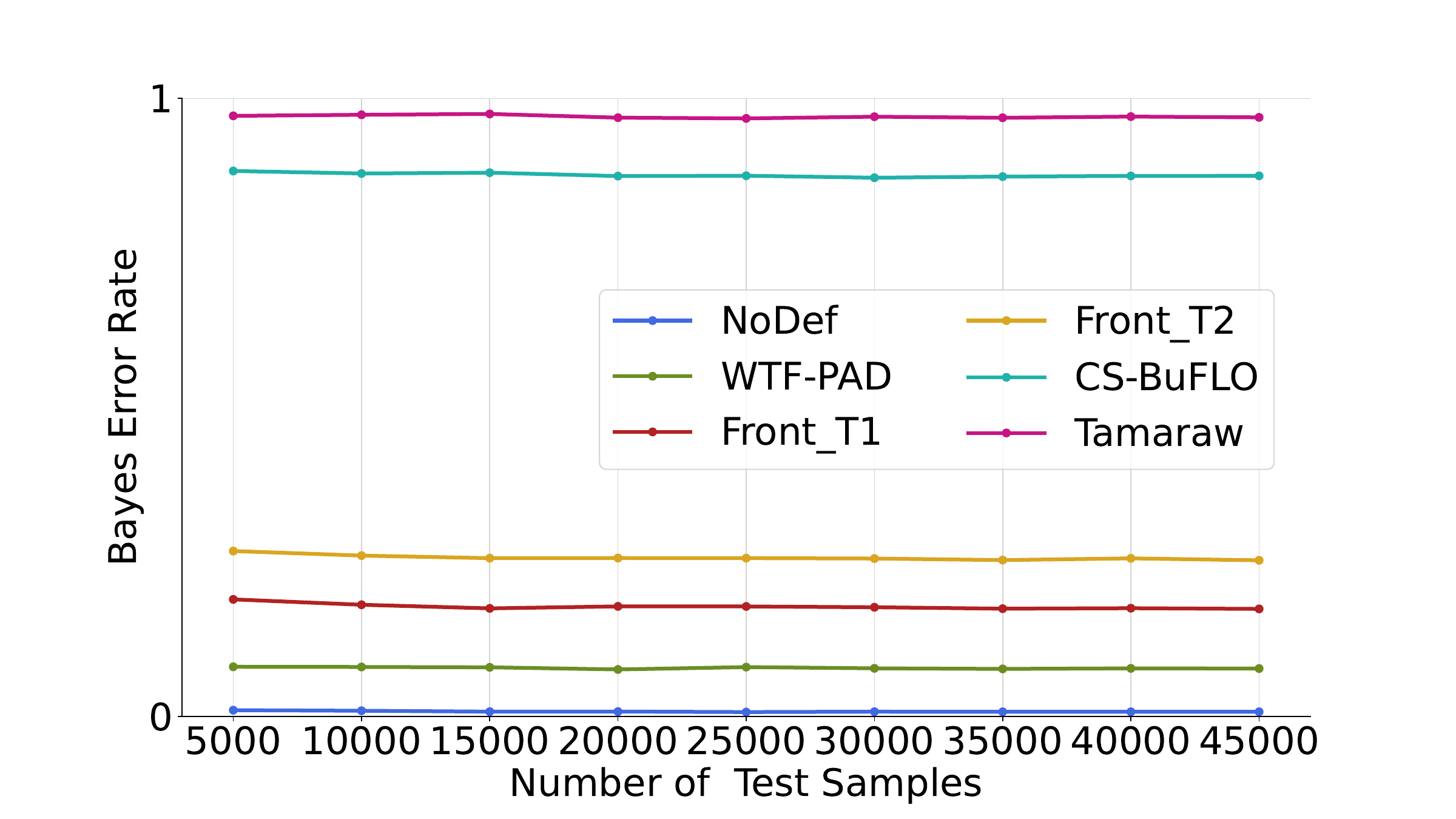}
        \vspace{-0.6cm}
        \caption{BER estimations for increasing number of samples for the kNN estimation while keeping the feature embeddings fixed (AWF$_{100}._{4500}$).}
        \label{fig:knn_convergence}
        \vspace{-0.3cm}
\end{figure}

\subsection{Convergence Plot for Mutual Information}
\label{sec:extended_convergence}

Figure \ref{fig:mi_convergence} shows the convergence for our MI estimation. We see that using more data increases the MI estimation, except for Tamaraw which decreases to 0.  Figure \ref{fig:knn_convergence} shows the convergence of the BER estimation as we increase the number of samples for the kNN estimation while keeping the feature embeddings fixed. The results show that our kNN-based BER estimator does not provide significantly different results with increasing amounts of data. This suggests that the major improvements on BER estimations when using more data (like those shown when comparing Figure~\ref{fig:estimationBoundComp} and Figure~\ref{fig:estimationScaleComp} are likely explained by the fact that the neural network is able to produce more accurate learned features.

\begin{table*}[t]
    \centering
    \vspace{1.3cm}
    \caption{DeepSE-WF Bayes error rate estimation (in \%) for multiple number of traces per website.}
    \vspace{-0.3cm}
    \begin{tabular}{lrrrrrrrr}
    \toprule
    \textbf{Traces}    & \textbf{60}             & \textbf{90}             & \textbf{100}            & \textbf{500}            & \textbf{1000}           & \textbf{2000}           & \textbf{4500} \\ \midrule
    NoDef     &  7.7 $\pm$ 0.6 &  5.1 $\pm$ 0.4 &  5.0 $\pm$ 0.3 &  1.4 $\pm$ 0.2 &  1.0 $\pm$ 0.0 &  0.6 $\pm$ 0.1 &  0.4 $\pm$ 0.0 \\
    WTF-PAD   & 34.0 $\pm$ 0.6 & 26.3 $\pm$ 1.2 & 24.3 $\pm$ 0.7 &  9.4 $\pm$ 0.4 &  7.0 $\pm$ 0.1 &  5.0 $\pm$ 0.3 &  4.2 $\pm$ 0.1 \\
    Front\_T1 & 54.5 $\pm$ 1.5 & 48.4 $\pm$ 0.7 & 46.0 $\pm$ 0.4 & 19.5 $\pm$ 0.5 & 15.6 $\pm$ 0.3 & 11.8 $\pm$ 0.6 &  9.9 $\pm$ 0.2 \\
    Front\_T2 & 61.2 $\pm$ 2.3 & 56.1 $\pm$ 1.2 & 53.6 $\pm$ 1.3 & 27.6 $\pm$ 0.3 & 22.7 $\pm$ 0.6 & 18.3 $\pm$ 0.2 & 14.8 $\pm$ 0.2 \\
    CS-BuFLO  & 69.6 $\pm$ 1.4 & 68.6 $\pm$ 0.7 & 68.7 $\pm$ 0.7 & 68.5 $\pm$ 0.3 & 67.9 $\pm$ 0.2 & 67.8 $\pm$ 0.1 & 67.2 $\pm$ 0.1 \\
    Tamaraw   & 84.1 $\pm$ 1.3 & 83.5 $\pm$ 1.5 & 83.9 $\pm$ 0.6 & 84.6 $\pm$ 0.9 & 84.4 $\pm$ 0.4 & 85.4 $\pm$ 0.5 & 85.4 $\pm$ 1.1 \\ \bottomrule
    \end{tabular}
    \label{tab:full_ber}
\end{table*}

\begin{table*}[t]
    \centering
    \vspace{1.2cm}
    \caption{DeepSE-WF mutual information estimation (in bits) for multiple number of traces per website.}
    \vspace{-0.3cm}
    \begin{tabular}{lrrrrrrrr}
    \toprule
    \textbf{Traces}    & \textbf{60}             & \textbf{90}             & \textbf{100}            & \textbf{500}            & \textbf{1000}           & \textbf{2000}           & \textbf{4500} \\ \midrule
    NoDef     & 5.93 $\pm$ 0.0 & 6.31 $\pm$ 0.0 & 6.34 $\pm$ 0.0 & 6.51 $\pm$ 0.0 & 6.54 $\pm$ 0.0 & 6.58 $\pm$ 0.0 & 6.59 $\pm$ 0.0 \\
    WTF-PAD   & 4.00 $\pm$ 0.0 & 4.82 $\pm$ 0.1 & 5.00 $\pm$ 0.0 & 6.00 $\pm$ 0.0 & 6.17 $\pm$ 0.0 & 6.38 $\pm$ 0.1 & 6.36 $\pm$ 0.0 \\
    Front\_T1 & 2.74 $\pm$ 0.0 & 3.41 $\pm$ 0.1 & 3.59 $\pm$ 0.0 & 5.34 $\pm$ 0.0 & 5.62 $\pm$ 0.0 & 5.99 $\pm$ 0.1 & 5.98 $\pm$ 0.0 \\
    Front\_T2 & 2.20 $\pm$ 0.0 & 2.82 $\pm$ 0.0 & 2.94 $\pm$ 0.1 & 4.74 $\pm$ 0.0 & 5.08 $\pm$ 0.0 & 5.45 $\pm$ 0.1 & 5.59 $\pm$ 0.0 \\
    CS-BuFLO  & 2.09 $\pm$ 0.0 & 2.34 $\pm$ 0.0 & 2.35 $\pm$ 0.0 & 2.46 $\pm$ 0.0 & 2.49 $\pm$ 0.0 & 2.48 $\pm$ 0.0 & 2.51 $\pm$ 0.0 \\
    Tamaraw   & 1.06 $\pm$ 0.0 & 0.98 $\pm$ 0.0 & 0.93 $\pm$ 0.0 & 0.29 $\pm$ 0.0 & 0.13 $\pm$ 0.1 & 0.08 $\pm$ 0.1 & 0.00 $\pm$ 0.0 \\ \bottomrule
    \end{tabular}
    \label{tab:full_mi}
\end{table*}

\begin{table*}[t]
    \centering
    \vspace{1.2cm}
    \caption{WFES Bayes error rate estimations (in \%) using the feature sets from WF attacks that use manually-crafted features over the AWF$_{100}._{90}$ dataset. Estimates marked in bold are used in Figure~\ref{fig:estimationBoundComp} and Figure~\ref{fig:estimationScaleComp}.}
    \vspace{-0.3cm}
    \begin{tabular}{lcccccc}
    \toprule
    \textbf{Attack Feature Set}   & \textbf{k-NN \cite{knnAttack}} & \textbf{k-FP \cite{kfingerprinting}} & \textbf{CUMUL \cite{cumul}} & \textbf{LL \cite{LLattack}} & \textbf{vng++ \cite{vng}} \\ \midrule
    NoDef     & 23.6 & 26.4 & \textbf{11.5} & 38.1 & 33.9 \\
    WTF-PAD   & 55.2 & \textbf{47.0} & 48.2 & 67.8 & 62.6 \\
    Front\_T1 & 71.2 & \textbf{66.9} & 71.9 & 80.8 & 78.7 \\
    Front\_T2 & 77.0 & \textbf{72.6} & 76.7 & 83.8 & 81.3 \\
    CS-BuFLO  & 68.0 & \textbf{67.5} & 67.8 & 71.0 & 67.8 \\
    Tamaraw   & 83.9 & 84.6 & 83.8 & 85.1 & \textbf{83.6} \\ \bottomrule
    \end{tabular}
    \label{tab:full_wfes}
\end{table*}

\begin{table*}[th!]
    \centering
    \vspace{1.3cm}
    \caption{Mutual information estimation (in bits) of WeFDE and MINE when using the AWF$_{100}._{4500}$ dataset.}
    \vspace{-0.3cm}
    \begin{tabularx}{0.82\linewidth}{XXX}
        \toprule
        \textbf{MI Estimator} & \textbf{WeFDE}     & \textbf{MINE}\\ \midrule
        NoDef            & 6.63 & 6.44 \\
        WTF-PAD          & 6.60 & 6.30 \\
        Front\_T1        & 6.59 & 6.01 \\
        Front\_T2        & 6.64 & 5.79 \\
        CS-BuFLO         & 6.55 & 2.53 \\
        Tamaraw          & 6.58 & 0.98 \\  \bottomrule
    \end{tabularx}
    \label{table:MI_results}
\end{table*}

\begin{figure*}[t]
    \centering
    \begin{subfigure}[t]{0.33\textwidth}
        \includegraphics[trim={0 0 0 2.4cm},clip,width=\textwidth]{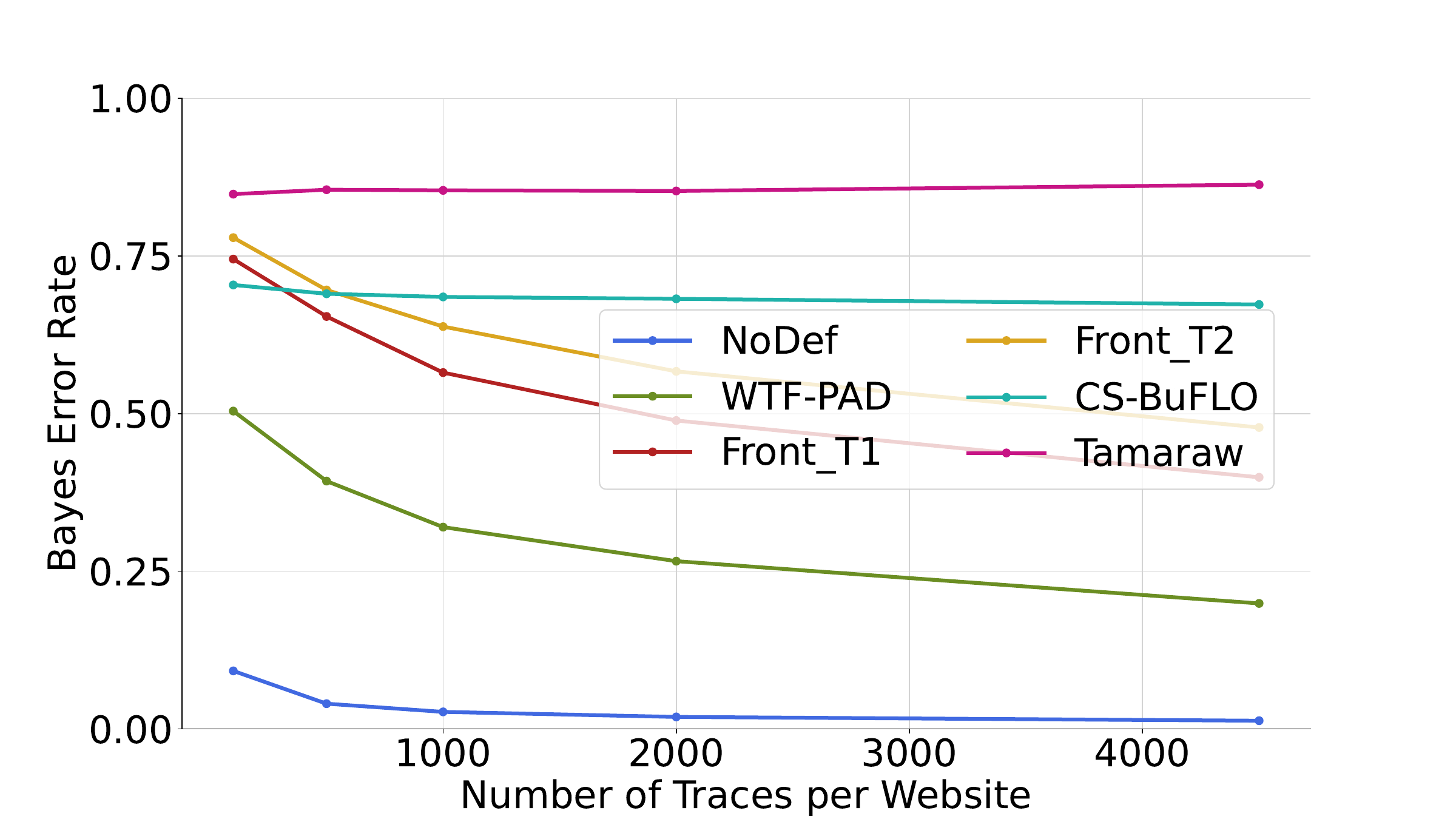}
        \vspace{-0.6cm}
        \caption{BER using AWF-CNN (AWF$_{100}._{4500}$).}
        \label{fig:convergence_awf}
    \end{subfigure}
    \begin{subfigure}[t]{0.33\textwidth}
        \includegraphics[trim={0 0 0 2.1cm},clip,width=\textwidth]{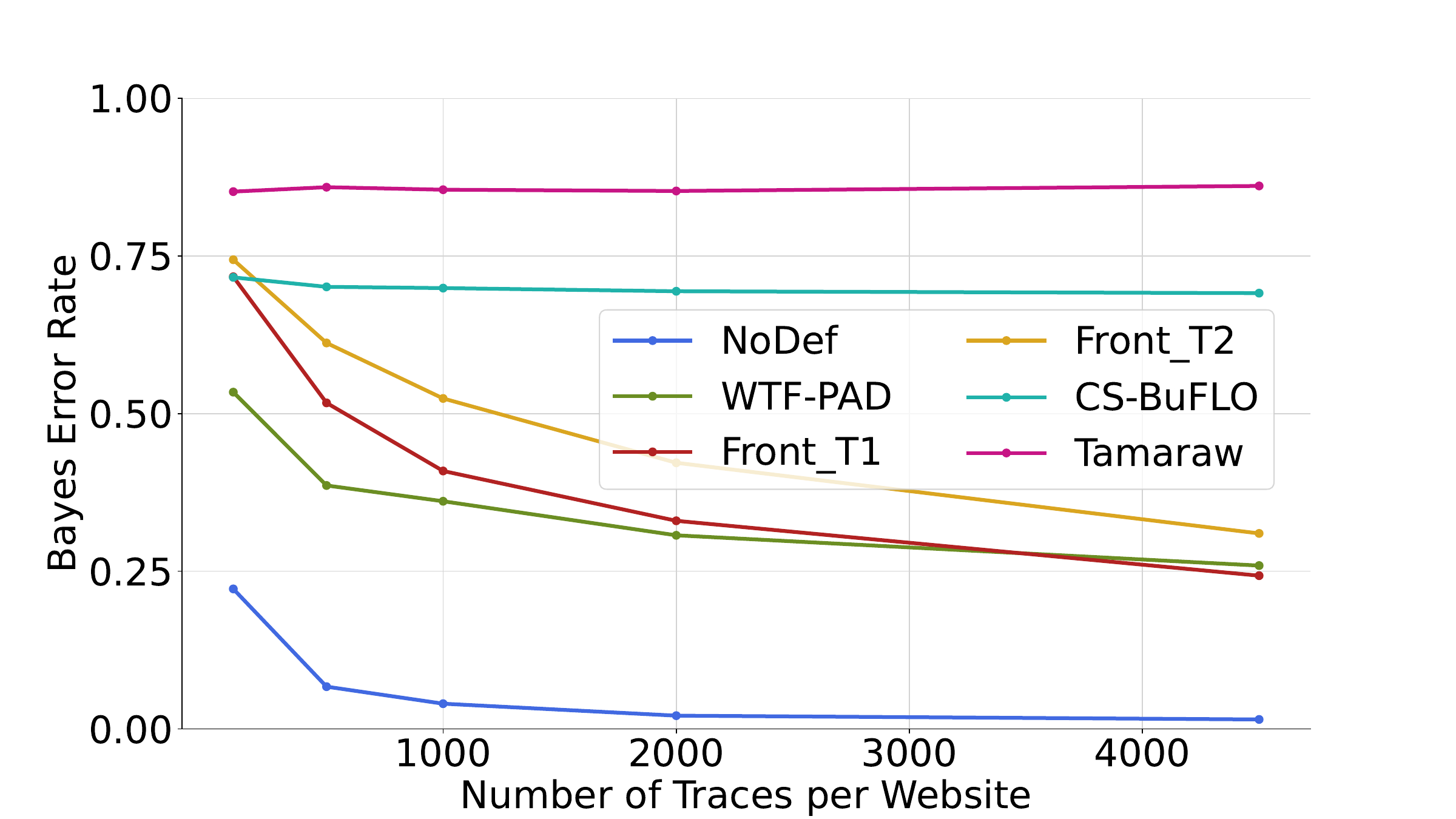}
        \vspace{-0.6cm}
        \caption{BER using TF - L2 loss (AWF$_{100}._{4500}$).}
        \label{fig:convergence_tf}
    \end{subfigure}
    \begin{subfigure}[t]{0.33\textwidth}
        \includegraphics[trim={0 0 0 2.4cm},clip,width=\textwidth]{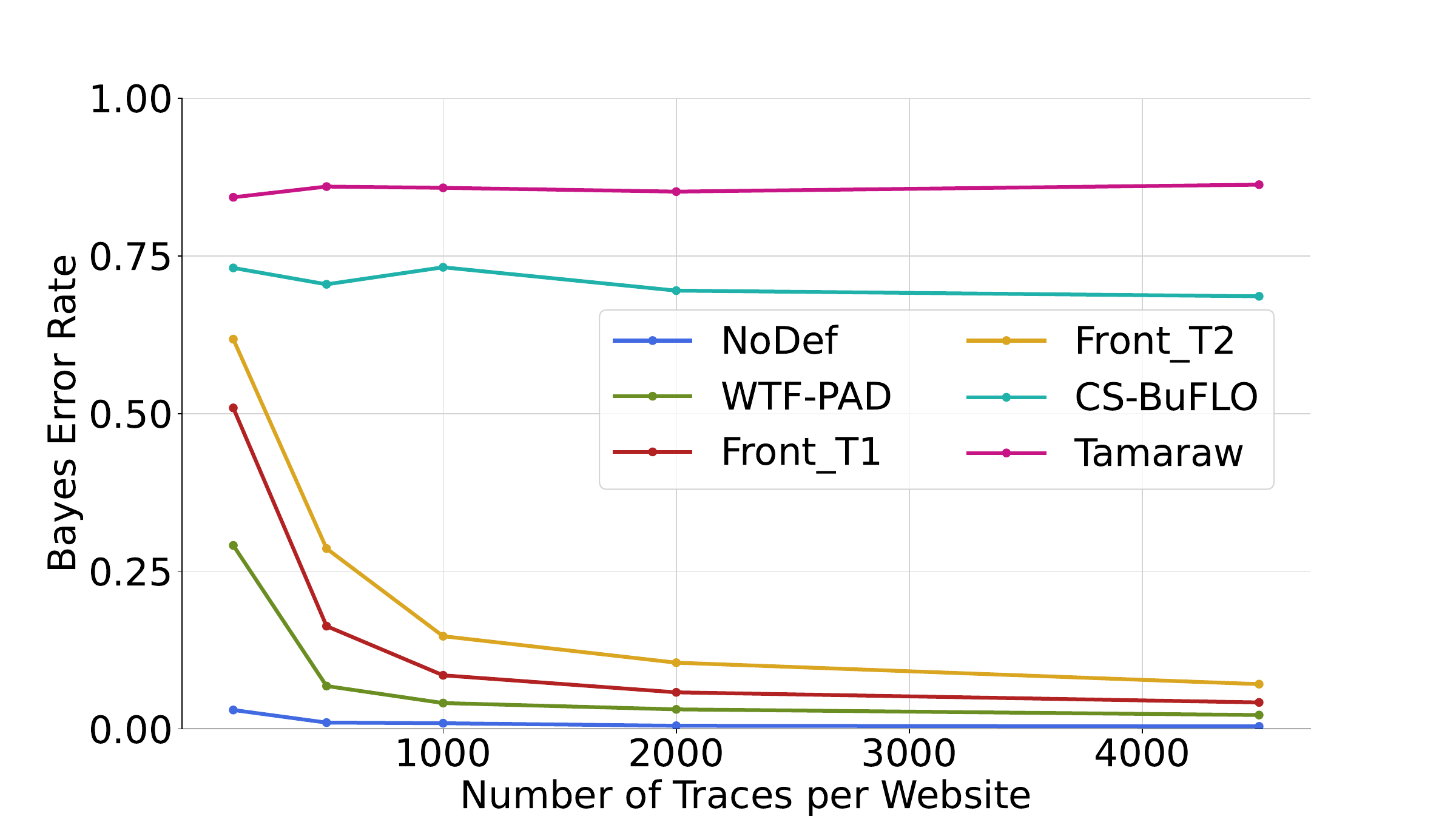}
        \vspace{-0.6cm}
        \caption{BER using Var-CNN (AWF$_{100}._{4500}$).}
        \label{fig:convergence_var}
    \end{subfigure}
    \vspace{-0.2cm}
     \caption{BER estimates obtained by DeepSE-WF when using AWF-CNN (a), TF - L2 loss 
 (b), and Var-CNN (c) architectures. 
     }
     \label{fig:architecture_convergence}
     \vspace{-0.2cm}
\end{figure*}

\subsection{Extended Analysis of TF Estimates}
\label{app:tf}

The main body of our paper presents the results for the TF DNN while using the L2 loss during the training process and L2 distance on the k-NN classification step. In this section, we shed additional light on the results obtained while using TF with the alternative cosine loss (and distance) also used in the original TF paper~\cite{triplet}.

\mypara{The TF attack error and DeepSE-WF's (TF) BER estimates.} Table~\ref{table:tf-awf4500} depicts the classification error of the TF attack and DeepSE-WF (TF)'s BER estimates on the AWF$_{100}._{4500}$ dataset, when using the L2 and cosine loss. We can see that, mirroring the conclusions from the original TF paper~\cite{triplet}, the use of cosine loss allows the TF attack to obtain better results than when using L2 loss. Specifically, while there is not a significant difference in the results obtained for constant rate defenses, the use of cosine loss in TF's training procedure allows the attack to decrease its classification error by as much as 17.3\% for the WTF-PAD defense.

We can observe a similar trend when analyzing the BER estimates produced by DeepSE-WF (TF), where the use of cosine loss in TF's training also allows for obtaining tighter estimates than those achieved when using L2 loss. For instance, for the Front\_T1 defense, DeepSE-WF (TF - L2 loss) achieves a BER of 24.3\%, while DeepSE-WF (TF - cosine loss) achieves a BER of 17.5\%. Nevertheless, DeepSE-WF's (TF - cosine loss) BER estimates are looser than those obtained when training DeepSE-WF with the DF or Var-CNN architectures (refer to Table~\ref{table:attackAndBoundEval}). As an example, while DeepSE-WF (TF - cosine loss) achieves a BER of 17.5\% for the Front\_T1 defense, DF achieves 9.9\% and Var-CNN achieves 4.2\%.

\mypara{TF experiments on the DS19 dataset.} Tables~\ref{table:tf-awf100} and~\ref{table:tf-ds19} depict the results obtained while using the two versions of TF to estimate the BER when using the AWF$_{100}._{100}$ and DS19$_{100}._{100}$ datasets. Both sets of experiments follow the trend previously identified in this section, where training TF with cosine loss allows us to obtain tighter BER estimates than when using L2 loss. Nevertheless, results still fall short of those obtained when using the DF or Var-CNN architectures (refer to Tables~\ref{table:AWFToDS19Results} and ~\ref{table:ds19Results}).

\subsection{Convergence Behavior and Efficiency of Alternative DNNs}
\label{app:dnn_scale}

In this section, we start by analysing the convergence behavior of other DNNs (AWF-CNN, TF - L2 loss, and Var-CNN) focused in Section~\ref{sec:dnn_arch_params} when producing security estimates with DeepSE-WF. Like our main results presented throughout Section~\ref{sec:evaluation}, we show the results obtained when training DeepSE-WF's DNNs according to the procedure described in Section~\ref{sec:dnn_arch_params}.
Then, we explore the efficiency of producing these BER estimates using the same DNNs.

\mypara{Convergence behavior:} 
Figure~\ref{fig:architecture_convergence} shows the BER estimates of DeepSE-WF using AWF-CNN, TF, and Var-CNN for an increasing number of traces per website. The plots show that all DNNs output a quite stable BER estimation for the constant-rate defenses, similarly to our observation when using the DF DNN (Figure~\ref{fig:ber_scaling_samples}). However, the plots also reveal that these architectures behave differently with respect to other defenses.
Compared to DF (Figure~\ref{fig:ber_scaling_samples}), AWF-CNN (Figure~\ref{fig:convergence_awf})) and TF (Figure~\ref{fig:convergence_tf})) show a less pronounced convergence for an increasing number of traces. For instance, for AWF-CNN, the BER estimation of Front\_T1 decreases only from 77.3\% to 39.9\%, whereas TF's BER estimation for Front\_T1 decreases from 71.7\% to 24.3\%.
In turn, Var-CNN (Figure~\ref{fig:convergence_var})) seems to struggle when the dataset size is small, with the BER estimate of Front\_T1 (67.1\%) being higher than the one for Front\_T2 (63.8\%) with AWF$_{100}._{100}$. The architecture shows large improvements on the estimate of the BER once the dataset size increases, with Front\_T1 decreasing from 67.1\% to 7.4\% when all traces are considered.

\begin{figure}[h]
    \centering
    \includegraphics[trim={0 0 0 2.4cm},clip,width=0.9\columnwidth]{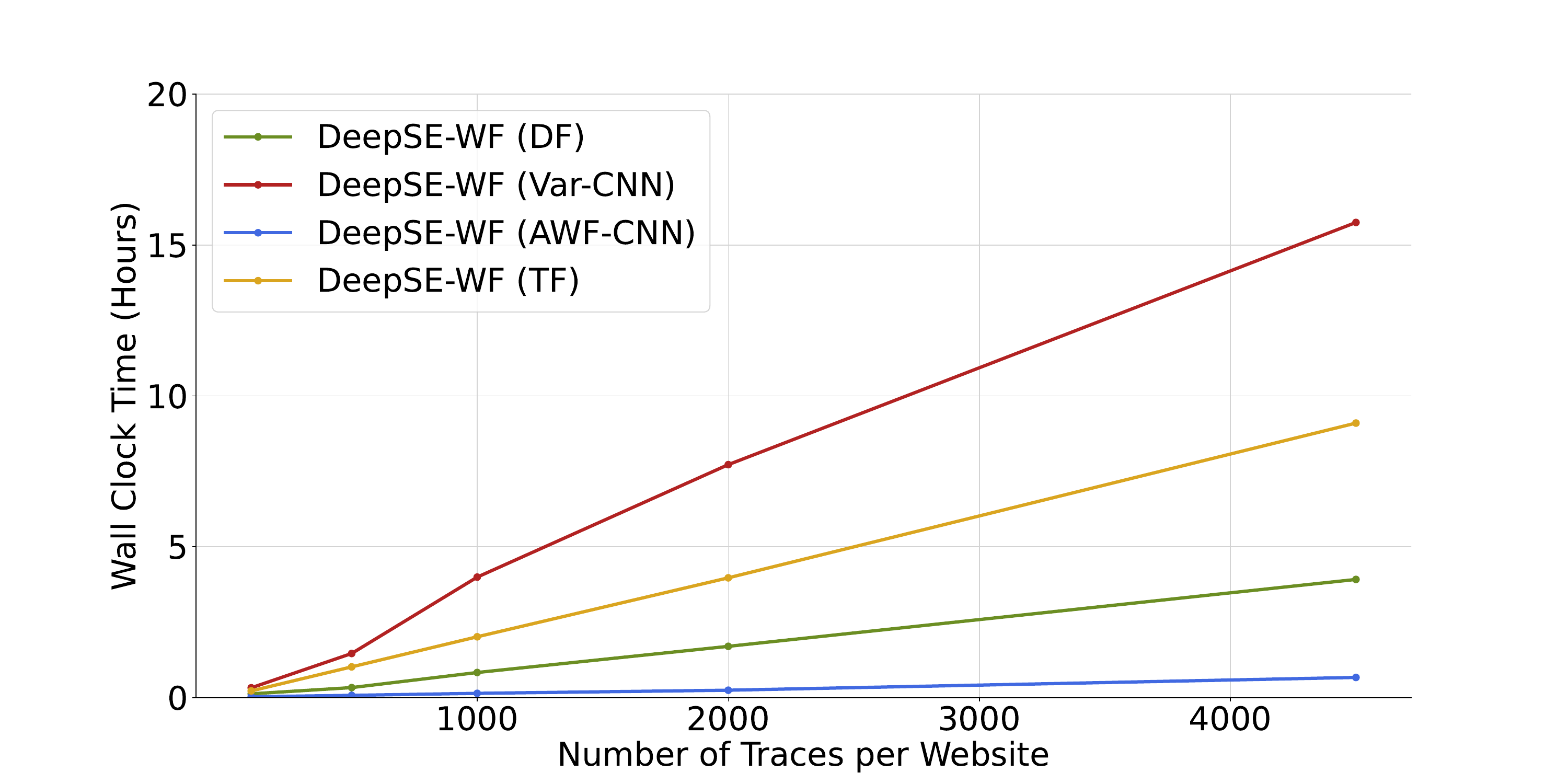}
    \vspace{-0.4cm}
    \caption{DeepSE-WF processing time for a single CV fold while increasing number of traces per website, while using different DNN architectures (AWF$_{100}._{4500}$).}
    \label{fig:processing_speed_deepse}
    \vspace{-0.55cm}
\end{figure}

\mypara{Efficiency:} 
Figure \ref{fig:processing_speed_deepse} depicts the processing time spent by each of the considered DNN architectures when producing estimates for different amounts of per-website samples. We can see that, while providing tighter BER estimates than other architectures, Var-CNN takes about 2, 4, and 20 times longer to train compared to TF, DF and AWF-CNN, respectively, on our GPU machine.

\begin{table}[t]
\centering
\caption{Classification error (in \%) for the TF attack on WF defenses compared to  TF BER estimates on AWF$_{100}._{4500}$.}
\vspace{-0.4cm}
\resizebox{\columnwidth}{!}{%
    \begin{tabular}{lcccccc}
        \toprule
        \textbf{Attacks \& Estimators}       & \textbf{NoDef}                & \textbf{WTF-PAD}                 & \textbf{Front\_T1}               & \textbf{Front\_T2}               & \textbf{CS-BuFLO}                & \textbf{Tamaraw} \\ \midrule
        TF (L2 loss) & 02.9 $\pm$ 0.4 & 45.4 $\pm$ 2.0 & 69.3 $\pm$ 0.7 & 77.5 $\pm$ 0.2 & 90.4 $\pm$ 0.2 & 97.5 $\pm$ 0.3 \\
        TF (cosine loss) & 03.0 $\pm$ 0.1 & 16.5 $\pm$ 0.2 & 31.9 $\pm$ 0.5 & 43.4 $\pm$ 0.1 & 89.9 $\pm$ 0.1 & 97.3 $\pm$ 0.3 \\
        \midrule
        DeepSE-WF (TF - L2 loss) & 01.5 $\pm$ 0.2 & 25.9 $\pm$ 1.3 & 24.3 $\pm$ 1.4 & 31.0 $\pm$ 3.7 & 69.1 $\pm$ 0.2 & 86.1 $\pm$ 1.2 \\
        DeepSE-WF (TF - cosine loss) & 01.0 $\pm$ 0.1 & 08.6 $\pm$ 0.1 & 17.5 $\pm$ 0.3 & 24.8 $\pm$ 0.1 & 69.0 $\pm$ 0.2 & 86.1 $\pm$ 1.3 \\
        \bottomrule
        \end{tabular}
}
    \label{table:tf-awf4500}
    \end{table}

\begin{table}[t]
\centering
\caption{BER estimates obtained by the two TF alternatives on the AWF$_{100}._{100}$ dataset.}
\vspace{-0.4cm}
\resizebox{\columnwidth}{!}{%
    \begin{tabular}{lcccccc}
        \toprule
        \textbf{Architecture}       & \textbf{NoDef}                & \textbf{WTF-PAD}                 & \textbf{Front\_T1}               & \textbf{Front\_T2}               & \textbf{CS-BuFLO}                & \textbf{Tamaraw} \\ \midrule
        DeepSE-WF (TF - L2 loss) & 25.3 $\pm$ 1.6 & 56.4 $\pm$ 1.8 & 73.0 $\pm$ 1.3 & 77.6 $\pm$ 0.9 & 71.3 $\pm$ 0.5 & 84.9 $\pm$ 0.9 \\
        DeepSE-WF (TF - cosine loss) & 17.7 $\pm$ 0.5 & 47.9 $\pm$ 0.8 & 65.8 $\pm$ 1.2 & 70.1 $\pm$ 1.1 & 71.0 $\pm$ 0.9 & 85.3 $\pm$ 0.7 \\
        \bottomrule
        \end{tabular}
}
    \label{table:tf-awf100}
    
\end{table}

\begin{table}[t]
\centering
\caption{BER estimates obtained by the two TF alternatives on the DS19$_{100}._{100}$ dataset.}
\vspace{-0.4cm}
\resizebox{\columnwidth}{!}{%
    \begin{tabular}{lcccccc}
        \toprule
        \textbf{Attacks \& Estimators}       & \textbf{NoDef}                & \textbf{WTF-PAD}                 & \textbf{Front\_T1}               & \textbf{Front\_T2}               & \textbf{CS-BuFLO}                & \textbf{Tamaraw} \\ \midrule
        DeepSE-WF (TF - L2 loss) & 12.7 $\pm$ 0.9 & 30.1 $\pm$ 0.5 & 61.6 $\pm$ 1.4 & 71.4 $\pm$ 1.7 & 67.9 $\pm$ 0.7 & 73.7 $\pm$ 1.3 \\
        DeepSE-WF (TF - cosine loss) & 08.3 $\pm$ 0.2 & 20.3 $\pm$ 0.5 & 42.4 $\pm$ 2.0 & 52.4 $\pm$ 1.6 & 67.9 $\pm$ 1.2 & 73.5 $\pm$ 1.3 \\
        \bottomrule
        \end{tabular}
}
    \label{table:tf-ds19}
    \vspace{-0.45cm}
\end{table}

\end{document}